\pdfoutput=1	
\documentclass[9pt,twocolumn,twoside]{pnas-new}
\setboolean{displaywatermark}{false}

\usepackage[T1]{fontenc}
\usepackage{textcomp}
\usepackage{booktabs} 
\usepackage{pifont}
\usepackage{amsfonts}
\usepackage{amssymb}
\usepackage{amsmath}
\usepackage{booktabs}
\usepackage{graphicx}
\usepackage{subfig}
\usepackage{xspace}
\usepackage{algpseudocode,algorithm,algorithmicx}

\MakeRobust{\Call}

\DeclareMathOperator*{\argmax}{argmax}

\newcommand{\SystemName}{DaltonQuant}

\newcommand{\avgLowRatio}{1.28}
\newcommand{\avgHighRatio}{1.40}
\newcommand{\avgLowPercent}{22\%}
\newcommand{\avgHighPercent}{29\%}

\newcommand{\SizeOfDataset}{28 million}
\newcommand{\MeanSpecimenAccuracy}{84\%}

\newcommand{\paletteSizeOne}{230}
\newcommand{\ratioLinearOnePng}{1.16}
\newcommand{\ratioNonLinearOnePng}{1.18}

\newcommand{\ratioLinearOneOrig}{3.15}
\newcommand{\ratioNonLinearOneOrig}{3.19}

\newcommand{\pctLinearOnePng}{13.9\%}
\newcommand{\pctNonLinearOnePng}{15.4\%}

\newcommand{\pctLinearOneTiny}{16.6\%}
\newcommand{\pctNonLinearOneTiny}{17.7\%}

\newcommand{\fileSizeOnePngQuant}{209,162}
\newcommand{\fileSizeOneLinearDalton}{179,995}
\newcommand{\fileSizeOneNonLinearDalton}{177,013}

\newcommand{\paletteSizeTwo}{204}
\newcommand{\ratioLinearTwoPng}{1.36}
\newcommand{\ratioNonLinearTwoPng}{1.29}

\newcommand{\ratioLinearTwoOrig}{2.84}
\newcommand{\ratioNonLinearTwoOrig}{2.69}

\newcommand{\pctLinearTwoPng}{26.4\%}
\newcommand{\pctNonLinearTwoPng}{22.2\%}

\newcommand{\pctLinearTwoTiny}{54.4\%}
\newcommand{\pctNonLinearTwoTiny}{48.5\%}

\newcommand{\fileSizeTwoPngQuant}{295,667}
\newcommand{\fileSizeTwoLinearDalton}{217,539}
\newcommand{\fileSizeTwoNonLinearDalton}{229,952}

\newcommand{\TinyPNG}{{\sc TinyPNG}\xspace}
\newcommand{\pngquant}{{\tt pngquant}\xspace}
\newcommand{\pngquantversion}{2.10.2 (August 2017)\xspace}

\newcommand{\lowHueAccuracyCorrelationLinear}{65\%}
\newcommand{\highHueAccuracyCorrelationLinear}{80\%}
\newcommand{\lowHueAccuracyCorrelationNonLinear}{37\%}
\newcommand{\highHueAccuracyCorrelationNonLinear}{66\%}
\newcommand{\lowDistanceReductionPct}{54\%}
\newcommand{\highDistanceReductionPct}{80\%}

\DeclareRobustCommand*{\deltau}{\ensuremath{\delta_{u}}}
\mathchardef\mhyphen="2D

\templatetype{pnasresearcharticle} 

\title{Incremental Color Quantization for Color-Vision-Deficient Observers Using Mobile Gaming Data}

\author[1]{Jos\'{e} Cambronero}
\author[2]{Phillip Stanley-Marbell}
\author[3]{Martin Rinard} 

\affil[1]{Computer Science and Artificial Intelligence Laboratory (CSAIL), MIT, Cambridge Massachusetts, MA 02139, USA}
\affil[2]{Department of Engineering, University of Cambridge, Cambridge CB3 0FA, UK}
\affil[3]{Computer Science and Artificial Intelligence Laboratory (CSAIL), MIT, Cambridge Massachusetts, MA 02139, USA}

\leadauthor{Cambronero} 

\authorcontributions{JC formulated the problem, designed the re-quantization algorithm, and performed the evaluation. All authors contributed to the writing.}
\correspondingauthor{\textsuperscript{1}To whom correspondence should be addressed. E-mail: {jcamsan@csail.mit.edu}}

\vspace{-0.1in}

\begin{abstract}
The sizes of compressed images depend on their spatial resolution
(number of pixels) and on their color resolution (number of color
quantization levels).  We introduce \SystemName{}, a new  color
quantization technique for image compression that cloud services
can apply to images destined for a specific user with known color
vision deficiencies. \SystemName{} improves compression in a
user-specific but reversible manner thereby improving a user's
network bandwidth and data storage efficiency.  \SystemName{}
quantizes image data to account for user-specific color perception
anomalies, using a new method for incremental color quantization
based on a large corpus of color vision acuity data obtained from
a popular mobile game.  Servers that host images can revert
\SystemName{}'s image re-quantization and compression when those
images must be transmitted to a different user, making the technique
practical to deploy on a large scale.  We evaluate \SystemName{}'s
compression performance on the Kodak PC reference image set and
show that it improves compression by an additional
\avgLowPercent{}--\avgHighPercent{} over the state-of-the-art
compressors \TinyPNG and \pngquant.
\end{abstract}

\dates{This manuscript was compiled on \today}

\begin{document}

\verticaladjustment{-5pt}

\maketitle
\thispagestyle{firststyle}
\ifthenelse{\boolean{shortarticle}}{\ifthenelse{\boolean{singlecolumn}}{\abscontentformatted}{\abscontent}}{}

\dropcap{I}mages account for a large fraction of the data transmitted on the

Internet~\cite{zettabyte}. The sizes of these images, when compressed
for storage and transmission, depend on their spatial resolution
(number of pixels) and color resolution (number of color quantization
levels). Existing image compression methods such as JPEG take into
account human visual perception but assume an individual with normal
color vision~\cite{wallace1992jpeg}.  Because compression algorithms
can provide improved compression when they target a smaller number
of distinguishable image colors, these algorithms could in principle
use information about color vision deficiencies in observers to
improve compression ratios.  Approximately 8\% of males and about
0.5\% of females have color vision deficiencies~\cite{byrne201011,
nie-colorblindness, xie2014color}. There is therefore a missed
opportunity to harness information about color vision deficiencies
to improve image compression and thereby to deliver improved network
speeds and improved file storage efficiency.

Users of mobile devices spend over 70\% of their time in applications
such as web browsing and productivity, as opposed to gaming
and multimedia~\cite{FlurryUiDominatesNotGames2017}. This observation, together with the personal
usage model of mobile devices and the popularity of user-specific
sites and web applications therefore makes it practical to transcode
images on a server to deliver improved performance customized to
individual users. Today, the fraction of almost 9\% of mobile
device owners with color vision deficiencies is missing out on the
opportunity for improved network performance and reduced data storage
requirements.

\subsection{\SystemName{}: Higher compression rates by exploiting human color vision deficiencies}
We introduce \SystemName{}, a new method for bespoke (user-specific)
image compression. \SystemName{} uses incremental color quantization
to improve image compression for observers with color vision
deficiencies, reducing file sizes by up to 29\% more than the state
of the art.  \SystemName{} builds a user-specific model for the
color acuity of its target users. This model characterizes the colors
which each target observer considers to be visually equivalent.  We
provide two different instantiations of \SystemName{}'s user-specific
function for quantifying color equivalences, constructed from a
dataset of \SizeOfDataset{} color comparisons performed by humans
on modern mobile displays. The dataset used was
collected through the mobile game \textit{Specimen}~\cite{specimen-game}.

We use an analysis of the data collected by the Specimen game,
across all its users, to identify individuals in the mobile game
data who demonstrated game behavior consistent with color vision
deficiencies. We use the data from 30 of these individuals to
construct and evaluate the two concrete instantiations of our
technique. We evaluate compression performance for these 30 individuals
and show that \SystemName{} enables average improvements of
\avgLowPercent{}\,--\,\avgHighPercent{} file size reduction (i.e.,
a compression ratio of \avgLowRatio{}\,--\,\avgHighRatio{}) over
state-of-the-art compressors \TinyPNG and \pngquant for the Kodak
PC image benchmark suite.  \SystemName{}'s improvements result from
undoing conservative decisions made by \TinyPNG and \pngquant,
neither of which customize compression for observers with color
vision deficiencies.

\textit{Color quantization} reduces the number of colors used to
represent an image and is a standard component of mainstream lossy
image compression techniques~\cite{deng1999peer, orchard1991color,
wu1992color}.  To avoid unacceptable image degradation, successful
color quantization techniques replace a large set of existing colors
with a smaller set that is perceived by most human observers to be
similar to the original set.  All existing color quantization
techniques assume a viewer with normal color vision and no vision
deficiencies when constructing their reduced color palette.  Because
an image compression algorithm may fail to combine colors that an
observer would not easily distinguish, this assumption leads to a
missed opportunity to provide improved compression ratios for color
vision deficient observers.

\subsection{The Specimen color matching data corpus}
We use data from Specimen~\cite{specimen-game}, an iOS game designed
to explore color perception.  Players of the Specimen game make
color comparisons by selecting colored objects ({\it specimens})
and matching them against a {\it target} background color
(Figure~\ref{fig:specimen-screenshot}). Specimen anonymously records each
of these color selection events using the Flurry~\cite{Yahoo:flurry}
cloud analytics service.  We use these anonymized per-user game
histories, which constitute \SizeOfDataset{} color comparisons as
of the time of writing, to determine which color pairs are consistently
confused by a specific player of the game. We then use these color
pairs as candidates for color mergers in the final quantized color
palette for image compression.  We are working with the developers
of the Specimen game to make both the data and our analysis tools
publicly available.

Although we take advantage of Specimen's large-scale dataset,
\SystemName{} is independent of the source of data on which colors
an observer consistently confuses. A mobile device user could choose
to quantify their color resolution capabilities through alternative
applications and \SystemName{} could use such data in its color
quantization algorithms.  As we show in our evaluation
(Section~\ref{sec:history}), the larger this color resolution dataset, the
better the color merger decisions and compression that \SystemName{}
can provide.

\subsection{Contributions}
In this work, we make the following contributions:
\begin{itemize}
  \item \textbf{\SystemName{} is the first compression method to
  purposefully exploit color vision deficiencies}: We introduce a
  new image compression color quantizer, \SystemName{}.  \SystemName{}
  is built on a user-specific function derived from mobile game
  data and a quantization algorithm that exploits properties of the
  user-specific function. The user-specific function quantifies
  colors which an observer will perceive to be equivalent (despite
  potential numeric differences)\footnote{We
  refer to these in the remainder of the paper as ``color confusions''.}.
  We implement two concrete instantiations of this function in
  \SystemName{} in order to evaluate our approach.  To the best of
  our knowledge, this is the first compression approach explicitly
  targeting users with color vision deficiencies and taking advantage
  of their differentiated color perception to reduce file sizes.

  \item \textbf{\SystemName{} outperforms production compressors}: 
  Our empirical evaluation of the \SystemName{} quantizer relative
  to popular PNG lossy compressors shows that \SystemName{} on average
  yields an additional file size reduction of
  \avgLowPercent{}--\avgHighPercent{}.  We validate our color
  confusion quantification functions and show that our proposed
  color mergers conform to expectations.
\end{itemize}

\section{Use Cases for Bespoke Image Compression}
Online services, such as Google Photos~\cite{googlephotos},
Flickr~\cite{flickr}, Dropbox~\cite{dropbox}, and Microsoft
OneDrive~\cite{msonedrive}, provide cloud-based storage for user
images. The file sizes for images affects storage requirements and
transfer performance on all of these server platforms as well as the
performance witnessed by mobile device clients. These services may use lossy
compression to reduce file sizes, subject to an acceptable threshold
on image quality degradation. For example, Flickr extensively
explored lossy compression of image
thumbnails: \textit{%
``The slight changes to image appearance went unnoticed by users, but
user interactions with Flickr became much faster,  especially for
users with slow connections, while our storage footprint became
much smaller.''%
}~\cite{flickrcompression}

\begin{figure}
\centering
\includegraphics[width=0.475\textwidth]{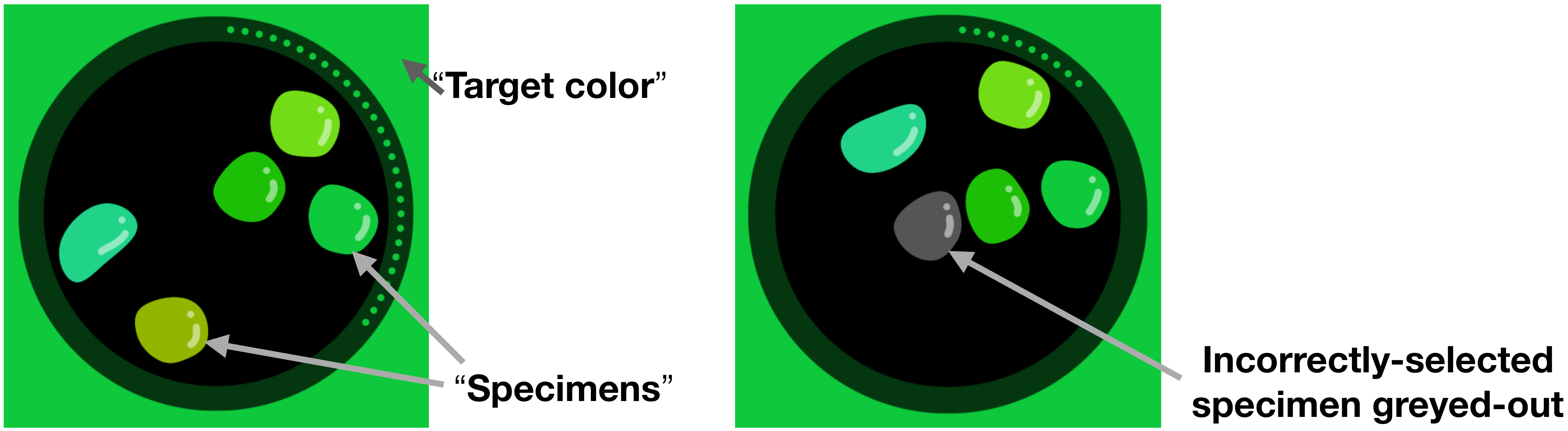}
\caption{An annotated screenshot of the Specimen game. The player's
goal is to select the \emph{specimen} that matches the \emph{target}
color. The application records event logs during game play to keep
track of correct and incorrect selections.}
\label{fig:specimen-screenshot}
\end{figure}

\SystemName{} produces a user-specific function
to quantify color confusions for color-vision-deficient users.
Compression algorithms can use this function to reduce image file size
beyond what is possible with state-of-the-art lossy PNG compressors
such as \TinyPNG and \pngquant. Any platform, such as Google Photos,
Flickr, or a personal mobile device, which constructs user profiles
and requires users to login to interact with their platform, can
store the \SystemName{} color confusion quantification function and
incorporate our technique to improve compression ratios.

For large-scale services, color-vision-deficient users can constitute
a relatively large population.  For example, as of May 2017 Google
Photos had 500 million monthly active users~\cite{googlephotousers}.
Given the accepted color blindness rates of 8\% in males and 0.5\%
in females~\cite{nie-colorblindness} and assuming equal gender
proportions in the user population implies up to 21 million users
of Google Photos could have their server-side data significantly
compressed by our approach.  In addition to the storage benefits,
the image size reductions will translate directly into improved
download speeds for these users.

In this work, we consider two main use cases for our compression
approach, both of which assume a personalized datastore, whether
cloud-based or local such as a personal mobile device:
\begin{itemize}
  \item \textit{Storage}: 
  Compressed images, along with a de-quantization map, reduce the
  storage requirements for server-side images as well as for locally-stored images
  such as those on a mobile phone. \textbf{The de-quantization map allows us
  to undo the incremental quantization for sharing the images with
  color-vision-normal individuals.}

  \item \textit{Transfer}: 
  When constraints favor network transfer improvements over storage
  concerns, the original image can be stored and server-side
  compression performed prior to transfer to the color-vision-deficient
  user, thereby improving network performance.
\end{itemize}

\section{Specimen: A Mobile Game About Color}
\label{sec:specimen-game}
Specimen~\cite{specimen-game} is an iOS game designed to allow users
to explore color perception.  It was purposefully designed to be
fun to use to enable wider deployment. Its large user base
and extensive game histories make it an appealing post-hoc data
source for research exploring color perception. We use data from 
users of Specimen to construct user-specific
functions which quantify colors that are numerically distinct but
which a given user cannot distinguish visually.  These functions
allow us to systematically reduce (i.e., re-quantize) the color palette
of a PNG image for compression while reducing the likelihood that
the target observer will notice image quality degradation.

A timer, indicated by the dots on the periphery of the black circular
region in Figure~\ref{fig:specimen-screenshot}, places an upper bound on the duration available for a color matching decision in the game.
If a player of the game incorrectly selects a specimen, the game
greys out the specimen, preventing it from being subsequently
re-selected.  If the correct specimen is selected, it disappears
and increases the player's score. Players advance through the game
by correctly matching colors in this manner. As players advance
through the game, the specimen and target colors presented to them
by the game become increasingly difficult to distinguish (as observed
by a color-vision-normal individual).

\section{Identifying Color-Vision-Deficient Users}
\label{sec:pcvd}
Color-vision-deficient individuals perceive parts of the visible
spectrum differently from color-vision-normal individuals.  These
differences can make it harder for color-vision-deficient individuals
to distinguish between two colors which color-vision-normal individuals
would perceive as distinct. To identify users that are potentially
color-vision-deficient, we define a heuristic based on decisions
that we believe correlate with color-vision-deficient users playing
Specimen.

Let $i$ be the $i$th turn in the history of Specimen selections
made by some user.  Let $Y_i$ be the target color in RGB space
observed by the user during turn $i$, and $\hat{Y}_i$ be the specimen
color in RGB space selected by that user in the same turn. Let
\textsc{Hue} define a mapping from RGB to the Hue dimension in the
Munsell Color System~\cite{munsell1907color}, \textsc{Lab}
define a mapping from RGB to CIE-LAB color space, and \textsc{Dist}
compute Euclidean distance.  Then we define the heuristic, $\mathcal{H}$, as:

{
\small
\begin{align}
\label{eqn:heuristic}
\mathcal{H} = \frac{\beta}{n}  \sum\limits_{i}^n \Call{Hue}{Y_i} \not = \Call{Hue}{\hat{Y_i}} + \frac{(1 - \beta)}{n}  \sum\limits_{i}^n  \Call{Dist}{\Call{Lab}{Y_i}, \Call{Lab}{\hat{Y_i}}} .
\end{align}
}

The heuristic $\mathcal{H}$ is a weighted sum of: \ding{202} the
fraction of color selections where the specimen ($\hat{Y_i}$) and
target ($Y_i$) colors were in distinct Munsell hues; \ding{203} the
average distance in the CIE-LAB color space, for all color selections.
In our experiments we set the combining weight $\beta$ to 0.5, as
this balanced both portions of the heuristic and retrieved multiple
individuals for evaluation.

We compute the score $\mathcal{H}$ for players of the Specimen game
who have made at least 1000 color comparisons (i.e., specimen
selections).  We rank the users in descending order using $\mathcal{H}$
and select the top 30 users for evaluation. We refer to these users
as {\it possibly color vision deficient (PCVD)} and refer to them
as \textit{PCVD User 1} through \textit{PCVD User 30} in the following
sections.

\begin{figure}
\centering
\includegraphics[width=0.475\textwidth]{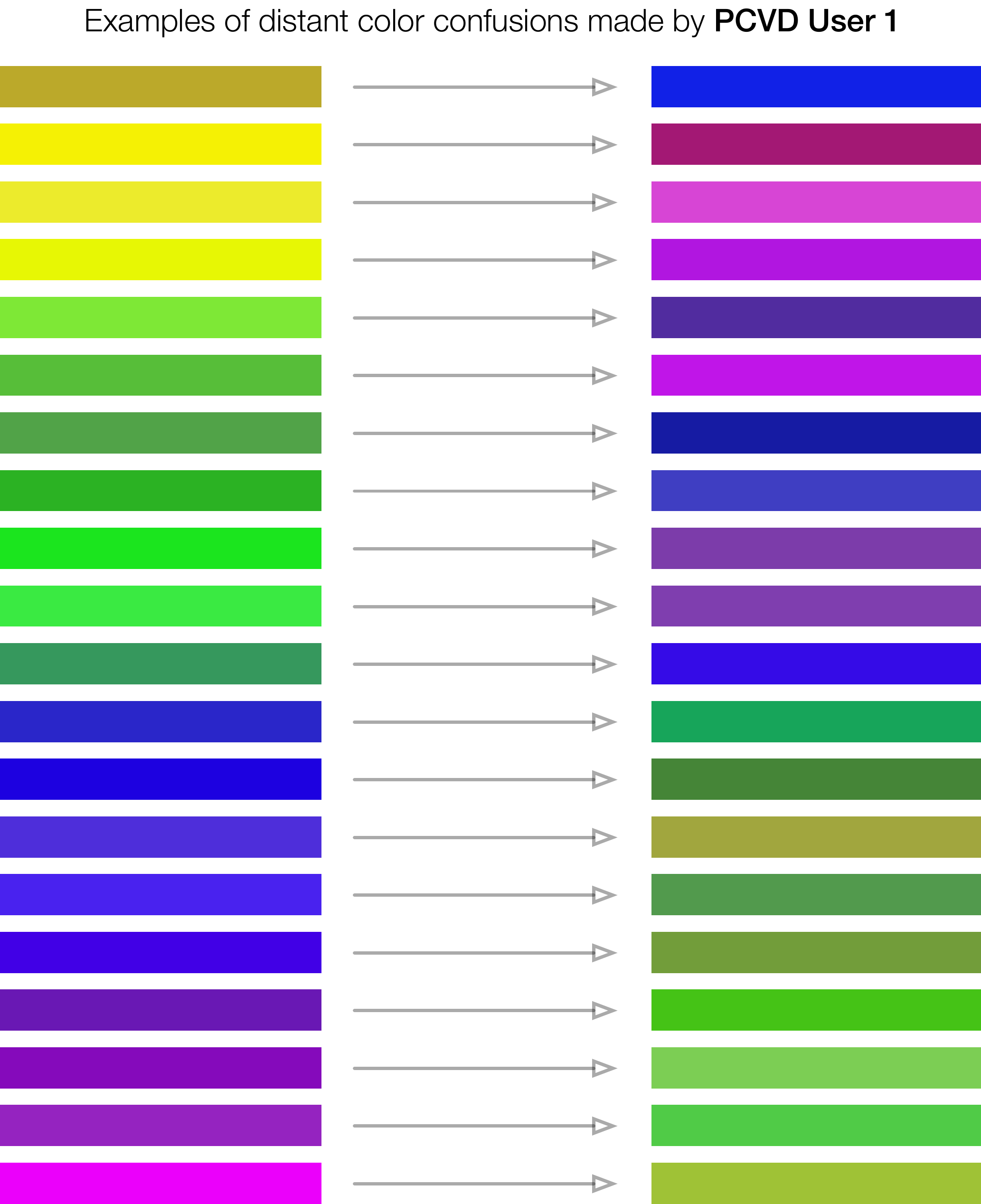}
\caption{
Color confusions made by the highest-ranking user (PCVD User 1 in
Figure~\ref{fig:cvd-user-accuracy}), based on the heuristic $\mathcal{H}$,
which was designed to identify color-vision-deficient users.  Repeated
confusions across hues support our hypothesis that our heuristic
successfully identified a set of potentially-color-vision-deficient
(PCVD) users.
}
\label{fig:confusion-samples}
\vspace{0.2in}

\centering
\includegraphics[scale=0.475]{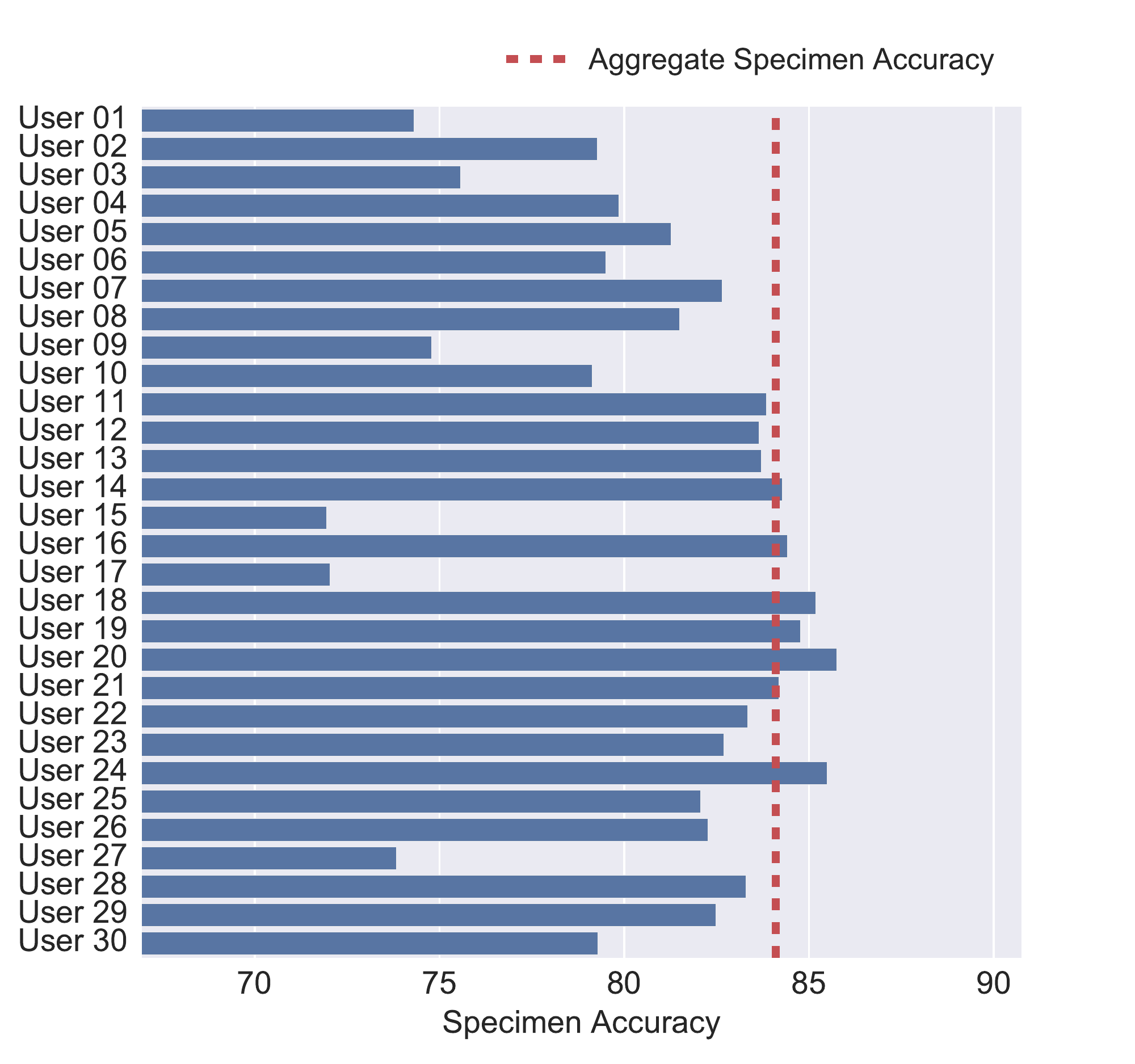}
\caption{
  Each blue bar corresponds to the selection accuracy in the
  Specimen game for a particular user. We present the top 30
  potentially color vision deficient (PCVD) users identified with
  our heuristic $\mathcal{H}$. Users are ordered based on the
  magnitude of $\mathcal{H}$. 24 of the 30 individuals used for
  evaluation have a Specimen
selection accuracy lower than the aggregate selection accuracy
of \MeanSpecimenAccuracy{} across all players.
}
\label{fig:cvd-user-accuracy}
\end{figure}

We validated the chosen users by visually inspecting a sample of
their color confusions, focusing on color choices where the target
and specimen colors correspond to different Munsell hue buckets.
Figure~\ref{fig:confusion-samples} shows the color confusions for the
highest ranking user based on the heuristic $\mathcal{H}$.  The
average Specimen selection accuracy across the 30 PCVD users is
80.9\% (4.11\% standard deviation) compared to the aggregate selection
accuracy of \MeanSpecimenAccuracy{} observed in the overall Specimen
dataset.  Figure~\ref{fig:cvd-user-accuracy} shows the selection accuracy
in the Specimen game for each of the 30 individuals along with the
aggregate selection accuracy in the dataset.

\section{Color Quantization With PCVD Data}\label{sec:quantization}
We present our approach for incremental image quantization based
on the color selection data for PCVD Users extracted from the Specimen Game.

\subsection{$\deltau$: A Bespoke Function for Quantifying Color Equivalence}\label{sec:delta-u-abstract}
We define a new user-specific function, $\delta_u$, that is a
continuous measure of color equivalence.  We use this function to
implement a color quantization algorithm that purposefully exploits
a user's inability to visually distinguish between certain pairs
of colors that are numerically distinct.  We first introduce an
abstract definition for the function $\delta_u$ below then we provide
two possible concrete definitions in 
Section~\ref{sec:quantization}.\ref{sec:delta-u-defs}
for evaluation of our approach.

We build a user's $\delta_u$ function from that user's history of
color confusions. The function takes two colors in the display's
native color space, usually the 24-bit RGB color space, and produces
a real value that correlates with the user's color confusions.
A user's $\delta_u$ will score a pair of colors higher the more likely that
pair is to be confused by the user. We use $\delta_u$ to guide the search for
colors that may be combined in an image palette during compression.

After constructing $\delta_u$ for a given user, we can take an
already quantized image and further reduce the image's
color palette.

\subsection{Color Quantization Optimization Formulation}
We formulate the problem of incremental quantization as an optimization
problem over two goals: \ding{202} increasing the number of pixels that are
modified to a given color in the output image and \ding{203} merging colors
that are more likely to be perceived as equivalent by the user. The
first goal serves as a proxy for the more complex goal of
merging neighboring pixels, which lend themselves to better encodings.
The second goal exploits color perception to remove the
visible effects of more aggressive color mergers.

We define a function $f_{\mathrm{color}}$ that combines these two
goals by using a weight, $\alpha$.  Let $c_{(\cdot)}$ stand for
different colors in the image palette $P$. Let $\mathit{pixels}(c_i)$
count the number of pixels with color $c_i$ in the original image.

\begin{align}
\label{eqn:combinedGoal}
\begin{split}
  f_{\mathrm{color}}(\alpha, c_i, c_j, \delta_u) = \\
  \alpha \frac{\delta_u(c_i, c_j)}{\max_{c_k, c_l \in P} \delta_u(c_k, c_l)} +\\
   (1 - \alpha) \frac{\mathit{pixels(c_i)}}{\max_{c_h \in P} \mathit{pixels}(c_h)} .
\end{split}
\end{align}

Let $m$ be a function that maps colors in the original palette
to a smaller palette, representing the quantization decisions. 
Let $X_m$ be the set of colors that are removed from the palette
based on the quantization decisions in $m$.
Our optimization goal is then defined as

\begin{align}
\label{eqn:optimizationGoal}
\argmax_{m} \sum_{c_i \in X_m} f_{\mathrm{color}}(\alpha, c_i, m(c_i), \delta_u) .
\end{align}

We approximate the optimization solution by performing a greedy
search over a list of candidate mergers $((c_i, c_j) | c_i, c_j \in
P \land c_i \neq c_j)$, sorted using the value of $f_{\mathrm{color}}$.
We take candidate mergers until we have reduced the color palette
size to a pre-defined target size provided by the user. 
We then unify merged colors transitively, so that if
color $c_1$ is merged with $c_2$, and then $c_2$ is merged with
$c_3$, any pixel in the image that was originally colored by $c_1$
will be colored by $c_3$ in the final version. Algorithm~\ref{algo:cvd-quantize}
summarizes this procedure.

\begin{algorithm}
 \caption{
   Incrementally quantizing the input palette $P$
   using $\delta_u$ for a PCVD user given a target number of colors $n$ in the output palette and a weight $\alpha$ that combines the two 
   optimization goals.
   }
   
    \label{algo:cvd-quantize}
    \begin{algorithmic}[1] 
    \Procedure{Quantize}{$\delta_u, \alpha, n, \mathit{image}$}
      \State $\mathit{mergers} \gets \{ \}$
      \State $\mathit{palette} \gets \Call{Palette}{\mathit{image}}$ \Comment{pre-quantized palette}
      \State Sort  $(c_i, c_j) \in P \times P$  based on $f_{\mathit{color}}(\alpha, c_i, c_j, \delta_u)$ \label{algo:cvd-quantize-sort}
      \State Take top mergers such that  $|P \setminus \mathit{mergers}| = n$
	    \State $\mathit{mergers} \gets \Call{Transitive-Closure}{\mathit{mergers}}$
	    \State \Return $\Call{Recolor}{\mathit{image}, \mathit{mergers}}$
    \EndProcedure
    \end{algorithmic}
\end{algorithm}

\subsection{Our Color Quantization Optimization is Reversible}
When performing the color changes, we maintain a map from the
original color to the pixel location for any pixels that are
recolored. This map allows us to reconstruct the original, pre-quantized
input image. The reconstructed pre-quantized image facilitates use cases
such as image sharing with color-vision-normal individuals.

\subsection{Defining $\deltau$: Transformation-based Metrics}\label{sec:delta-u-defs}
We explore two different concrete definitions for $\delta_u$. We provide
these definitions as examples that satisfy our $\delta_u$ abstract
definition (presented previously in Section~\ref{sec:quantization}.\ref{sec:delta-u-abstract})
and provide effective compression, as shown in our empirical results,
which follow in Section~\ref{sec:results}. Other such definitions may
exist.

A possible approach to modeling user color vision deficiencies is
to perform a transformation that takes the standard CIE-LAB color
space and produces a user-customized space. Prior work~\cite{jefferson2007interface}
has used a transformation-based approach to perform tasks
such as user interface adaptation for color-vision-deficient
users, though no prior work has built this
transform using large-scale data collected through mobile devices
as we do in \SystemName{}. To implement this approach,
we first construct a matrix $Y$ where each row corresponds to a target
color, converted to $\mathrm{CIE \mhyphen LAB}$, observed by a
Specimen user. Similarly, we construct $\hat{Y}$ with the corresponding
selected colors. We can then build a transformation $f$ such that
$f(Y) \approx \hat{Y}$. Using $f$, colors that were perceived as
close by color-vision-deficient individuals, despite potentially
large distances in CIE-LAB space, will be closer in the transformed
space.

\subsubsection{A Linear Transformation}
We consider a simple linear transformation, represented as a matrix
$M_{3 \times 3}$ such that $Y M \approx \hat{Y}$. We minimize
$\left\lVert \hat{Y} - Y M \right\rVert^2$, which we can solve as
a simple ordinary-least-squares problem.

\subsubsection{A Non-Linear Transformation}
Rather than construct a set of basis functions that include non-linear transformations over $Y$, 
we can provide non-linearity by instantiating
$f$ to be an off-the-shelf neural network~\cite{scikit-learn}.
The default implementation we used has a single hidden layer of 100 units
and rectified linear unit (\textit{ReLU}) activation functions~\cite{nair2010rectified}.

\subsubsection{From Transformation to $\delta_u$}
For both the linear and non-linear transformations described above, we define $\delta_u$
to be the inverse of the Euclidean distance between the two input colors
in the transformed space. For the non-linear transform,
two distinct input colors may map to the same color in the transformed
space. We therefore add a small value $\epsilon$ to all distances to avoid
a degenerate division by zero.

\section{Why Incremental Quantization?}
Traditional color quantization techniques are usually designed to
merge colors that are near each other according to some chosen
distance function and color space.  The standard definition of
proximity of colors is their distance in one of the color spaces
such as the RGB color space or the CIE-LAB color space. RGB is the
native color space of most displays, but Euclidean distances in RGB
do not correspond to color differences perceived by human observers.
The CIE-LAB color space on the other hand was designed so that
Euclidean distances in CIE-LAB correspond to color differences
perceived by humans.

Median-cut color quantization~\cite{kruger1994median}, a popular
first step in quantization, sorts the color palette by the dimension
with the largest range, and then partitions and recursively merges
colors by averaging.  These techniques are fast and effective. But
if there are enough colors in the original palette, techniques such
as median-cut do not usually have the ability to merge colors that
are perceived as far away by color-vision-normal individuals but
as close by color-vision-deficient individuals.

Production-quality compressors also often make conservative decisions,
which may result in larger file sizes, in order to preserve perceived
image quality quality. For example, most common compressors will
employ error-diffusion techniques to eliminate artifacts that arise
through quantization.  These are reasonable tradeoffs when considering
color-vision-normal users, who might easily distinguish between
distinct reds or blues, for example. But these changes may be overly
conservative when considering a color-vision-deficient user.  For
example, a viewer with Protanopia might have a hard time distinguishing
shades of reds and shades of blue~\cite{ishihara1960tests}.

Our incremental color quantization exploits both near- and long-range
color confusions to reduce the color space and undo some of the
conservative choices that the original compressor may have made.
By starting with an image that has already been quantized by tools
such as \TinyPNG or \pngquant, we can increase the compression
ratios for color-vision-deficient users while easily reverting to
the original image (which is already compressed) for color-vision-normal
users.  This allows incremental server-side data compression and
network communication improvement for client devices for
color-vision-deficient users, without directly impacting
color-vision-normal users.

\section{Results}\label{sec:results}
We evaluate our approach on the Kodak PC Set~\cite{franzen1999kodak},
which consists of 24 pictures ranging from portraits to action shots
and landscapes. We applied \SystemName{} using 30 different Specimen
users as observers. We chose these 30 individuals as they had at
least 1000 game observations and ranked highest according to our
heuristic $\mathcal{H}$ (Equation~\ref{eqn:heuristic}) for PCVD
user identification.  For additional analyses, such as the impact
of different game histories, we take PCVD User 1 to be our prototypical
color-vision-deficient user as their game history produced the
highest value for $\mathcal{H}$.

\subsection{Methodology}
We constructed a $\delta_u$ color equivalence function using each
of the approaches described in Section~\ref{sec:quantization}.\ref{sec:delta-u-defs} for each of
the 30 PCVD users presented previously in Section~\ref{sec:pcvd}.  Our
experiments evaluate the additional compression provided by our
PCVD-specific color quantization and our results demonstrate the
benefits of our color quantization when compared to three other
quantizers: \pngquant, \TinyPNG, and basic median-cut color
quantization.

\pngquant is an open source PNG lossy compression utility. It uses
a modified iterative version of median-cut quantization along with
a perception model and performs color corrections on the final
palette~\cite{pngquant}. We used \pngquant version \pngquantversion,
on Mac OS X. \TinyPNG, created by Voormedia,  is a proprietary lossy
PNG compressor with a limited amount of free monthly compressions
and a subscription-based model for further requests~\cite{tinypng}.
The basic median-cut quantizer is a standard recursive implementation
of the median-cut algorithm for color quantization~\cite{kruger1994median}.

Both \pngquant and the median-cut quantizer can be given a target
number of colors for the final palette, while \TinyPNG does not
take additional arguments in its API.  For all evaluation, we
initially quantize the images using one of the three quantizers
introduced. Where possible the target number of colors for all
pre-quantization is set to 256.  We then choose a final target
number of colors from (230, 204, 179, 153, 128), which are approximately
10\%-spaced decreases from the original 256 colors, and apply
\SystemName{}'s incremental quantization algorithm to reach that
palette size.

We define \emph{the reference image} as follows.  For both \pngquant
and median-cut quantization, which allow us to specify the palette
size, the reference image has the same number of colors in its
palette as our output image. For \TinyPNG, which does not accept a
palette size parameter, we define the reference image to be the
compressed image it produces and which is used by our technique as
a starting point for incremental quantization. When discussing file
size reductions and compression ratios we give these metrics relative
to the reference image and the original image (uncompressed and
without any form of quantization) where relevant.

For all evaluations, we set the multi-objective optimization weighting
factor $\alpha$ (Equation~\ref{eqn:combinedGoal} and
Equation~\ref{eqn:optimizationGoal}) to 0.5, which represents an
equal weighting of the two optimization goals: maximizing the sum
of $\delta_u$ over the colors merged and maximizing the number of
pixels with a given color.

\subsection{Our PCVD Color Quantization Outperforms State-of-the-Art CVD-Agnostic Quantization}

\begin{figure*}
\centering
\subfloat[\textbf{Original.}]{\includegraphics[width=0.32\textwidth]{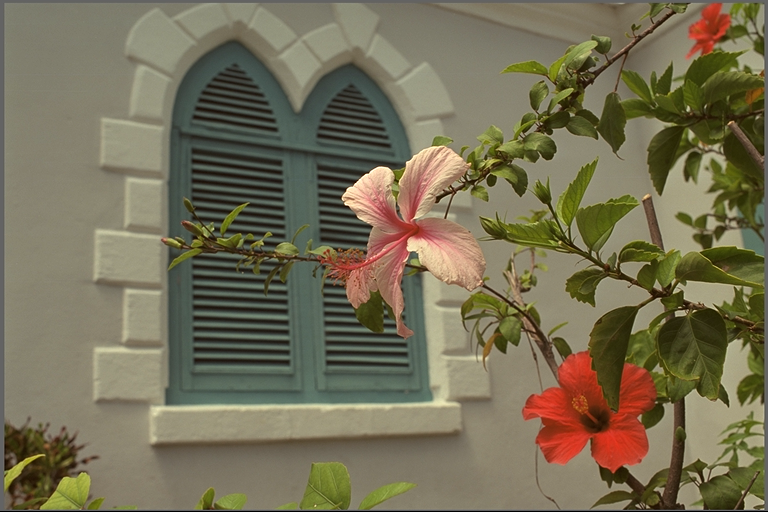}}~~~
\subfloat[\textbf{$\delta_u$: Linear transformation.}~  \SystemName{} produces an image
of \fileSizeOneLinearDalton{} bytes.  With the same palette size,
\pngquant{}'s image is \fileSizeOnePngQuant{}.  This is a compression
ratio of \ratioLinearOneOrig{} relative to the original image, and
\ratioLinearOnePng{} relative to the \pngquant{} image with the same
palette size.]{\includegraphics[width=0.32\textwidth]{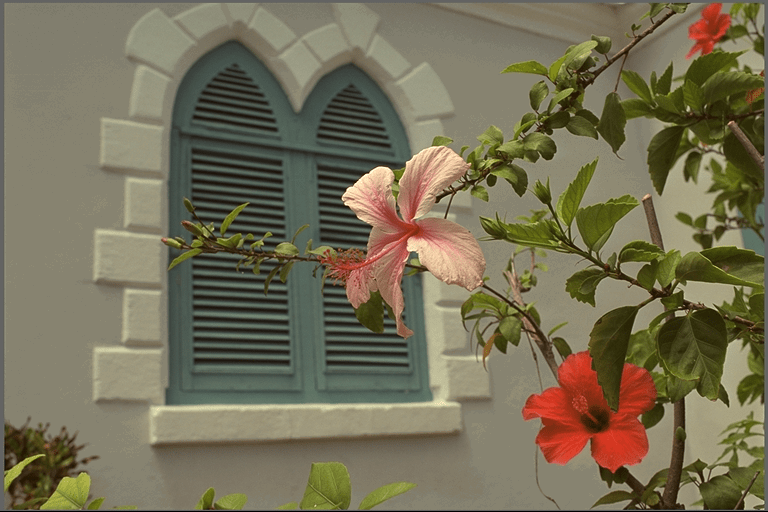}}~~~
\subfloat[\textbf{$\delta_u$: Non-linear transformation.}~  \SystemName{} produces an image
of \fileSizeOneNonLinearDalton{} bytes.  With the same palette size,
\pngquant{}'s image is \fileSizeOnePngQuant{}.  This is a compression
ratio of \ratioNonLinearOneOrig{} relative to the original image, and
\ratioNonLinearOnePng{} relative to the \pngquant{} image with the same
palette size.]{\includegraphics[width=0.32\textwidth]{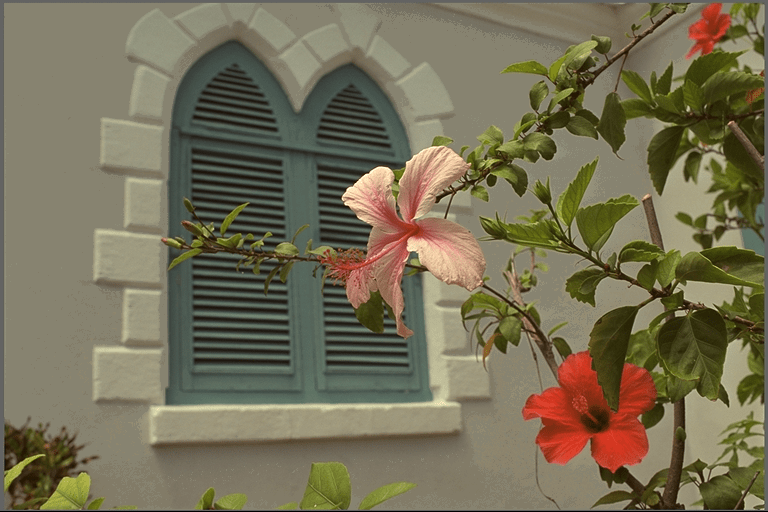}}
\caption{
The transformation-based $\delta_u$ implementations for PCVD user
1 with palette size of \paletteSizeOne{}.
}
\label{fig:compression-example1-transform}
\end{figure*}

Figure~\ref{fig:compression-example1-transform} presents the effects of
our incremental PCVD color quantization on image 7 of the Kodak
benchmark based on the $\delta_u$ functions constructed from PCVD User 1's
data.  Both $\delta_u$ implementations produced smaller files than the \pngquant{}
output with the same palette size. With a target palette size of
\paletteSizeOne{}, the linear and non-linear transformation-based
$\delta_u$ enable file size reductions of \pctLinearOnePng{} and
\pctNonLinearOnePng{}, respectively, relative to the reference image.
The file size reductions relative to \TinyPNG are \pctLinearOneTiny{}
and \pctNonLinearOneTiny{} for the linear and non-linear
transformation-based $\delta_u$, respectively. We note that,
when comparing to \TinyPNG, we cannot control the output palette
size for the reference image.

\begin{figure*}
\centering
\subfloat[\textbf{Original.}]{\includegraphics[width=0.32\textwidth]{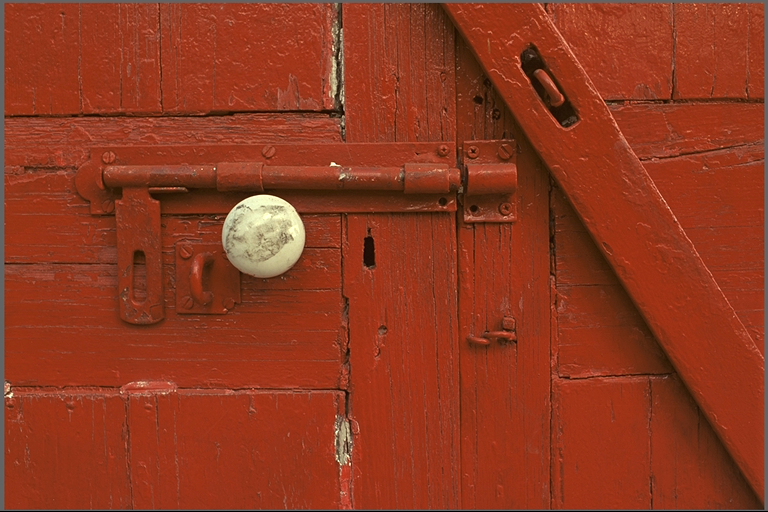}}~~~
\subfloat[\textbf{$\delta_u$: Linear transformation.}~  \SystemName{} produces an image
of \fileSizeTwoLinearDalton{} bytes.  With the same palette size,
\pngquant{}'s image is \fileSizeTwoPngQuant{}.  This is a compression
ratio of \ratioLinearTwoOrig{} relative to the original image, and
\ratioLinearTwoPng{} relative to the \pngquant image with the same
palette size.]{\includegraphics[width=0.32\textwidth]{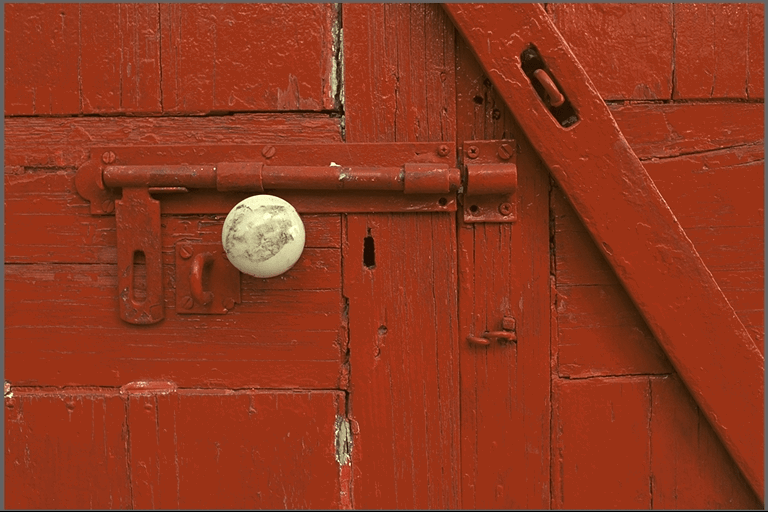}}~~~
\subfloat[\textbf{$\delta_u$: Non-linear transformation.}~  \SystemName{} produces an image
of \fileSizeTwoNonLinearDalton{} bytes.  With the same palette size,
\pngquant{}'s image is \fileSizeTwoPngQuant{}.  This is a compression
ratio of \ratioNonLinearTwoOrig{} relative to the original image, and
\ratioNonLinearTwoPng{} relative to the \pngquant image with the same
palette size.]{\includegraphics[width=0.32\textwidth]{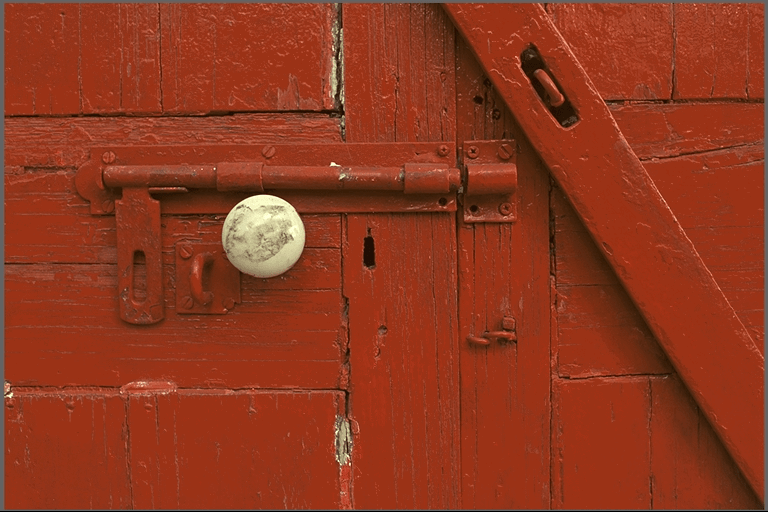}}\\
\caption{
The transformation-based $\delta_u$ implementations for PCVD user
1 with palette size of \paletteSizeTwo{}.
}
\label{fig:compression-example2-transform}
\end{figure*}

Figure~\ref{fig:compression-example2-transform} presents the effects of
our incremental PCVD color quantization on image 2 of the Kodak
PC image benchmarks, based on the $\delta_u$ functions constructed from PCVD User 1's
data.  Both implementations produced smaller files than the \pngquant{}
output with the same palette size. With a target palette size of
\paletteSizeTwo{}, the linear and non-linear transformation-based
$\delta_u$ enabled file size reductions of \pctLinearTwoPng{} and
\pctNonLinearTwoPng{} relative to the reference image, respectively.
The file size reductions relative to \TinyPNG are \pctLinearTwoTiny{}
and \pctNonLinearTwoTiny{} for the linear and non-linear
transformation-base $\delta_u$, respectively. Again, we note that,
when comparing to \TinyPNG, we cannot control the output palette
size for the reference image.  As expected, a smaller palette target
size results in larger file size reductions.

As the number of colors for the output image decreases, the file
size reductions achieved by \SystemName{} increases.  In contrast
to standard lossy compression, the artifacts that may arise are
designed, by the formulation of the color equivalence quantification
function $\delta_u$, to be less likely to be visible by PCVD users.
We provide evidence supporting this assertion in Section~\ref{sec:validation}.\ref{sec:validating-delta-u}.

\begin{figure*}
\centering
\includegraphics[width=1.0\textwidth]{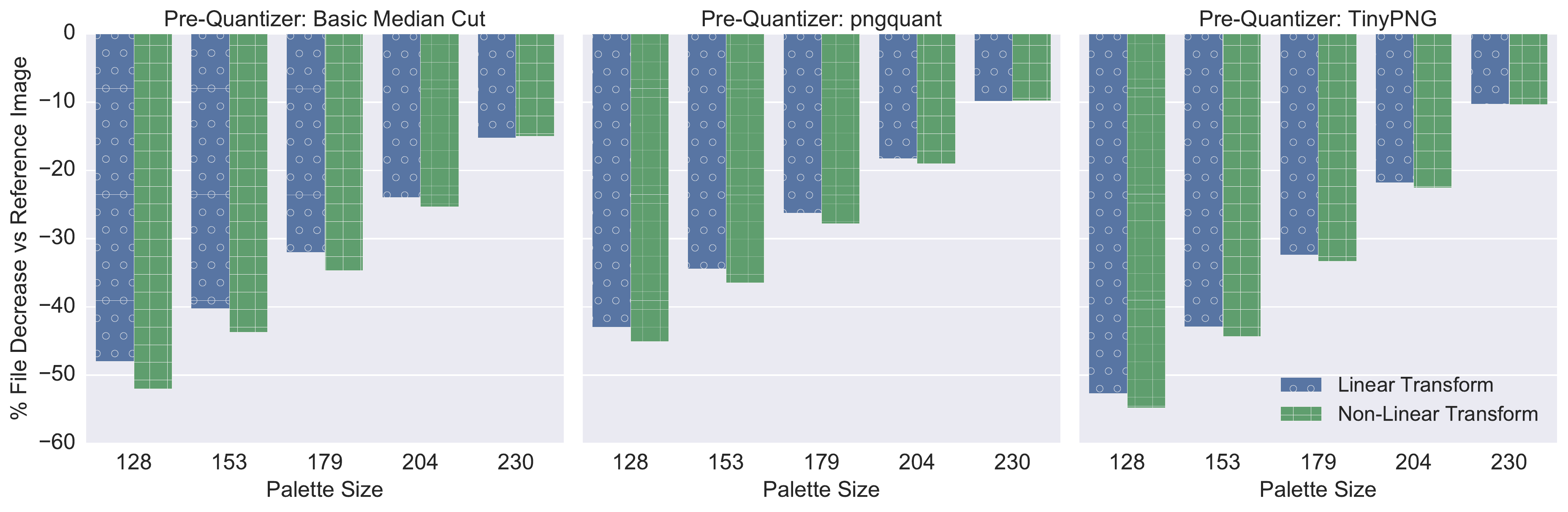}
\caption{
Incremental quantization yields file size decreases between \avgLowPercent{}
and \avgHighPercent{}, on average, compared to the reference image
file sizes with the same palette size. Smaller palettes lead to larger file size decreases.
Evaluation is done over the Kodak PC Set.
}
\label{fig:barplot-file-sizes}
\end{figure*}

Figure~\ref{fig:barplot-file-sizes} shows a summary of incremental file
size reductions achieved by applying $\delta_u$ color quantization
at several output palette sizes.  The data in Figure~\ref{fig:barplot-file-sizes}
are averages across the 30 PCVD users we selected for evaluation.
The file size reductions in Figure~\ref{fig:barplot-file-sizes} correspond
to the percentage change in file size from the reference image
to the image produced by \SystemName{}. A smaller palette
size correlates with a larger amount of lossy compression.  We also
observe that the non-linear transformation $\delta_u$
produces more significant file size reductions than the linear transformation $\delta_u$.

Table~\ref{table:mean-absolute-file-sizes} presents the average absolute
file sizes for the reference image and \SystemName{}'s output for
the Kodak PC image set under different palette sizes and using
different initial quantizers. For \pngquant and Basic Median-Cut
entries, the reference image and \SystemName{}'s output have color
palettes of the same size.

\begin{table}
\centering
\resizebox{\columnwidth}{!}{
\begin{tabular}{rlrr}
\toprule
 Palette Size &     Pre-Quantizer &  Reference (kB) &  DaltonQuant (kB) \\
\midrule
          128 &  Basic Median Cut &          168.51 &             84.56 \\
          128 &          pngquant &          198.90 &            112.31 \\
          128 &           TinyPNG &          213.03 &            100.10 \\
          153 &  Basic Median Cut &          176.43 &            102.97 \\
          153 &          pngquant &          209.97 &            136.51 \\
          153 &           TinyPNG &          213.03 &            121.47 \\
          179 &  Basic Median Cut &          184.66 &            123.64 \\
          179 &          pngquant &          220.64 &            162.09 \\
          179 &           TinyPNG &          213.03 &            144.23 \\
          204 &  Basic Median Cut &          191.99 &            145.36 \\
          204 &          pngquant &          229.65 &            187.81 \\
          204 &           TinyPNG &          213.03 &            166.72 \\
          230 &  Basic Median Cut &          200.33 &            170.56 \\
          230 &          pngquant &          238.65 &            215.78 \\
          230 &           TinyPNG &          213.03 &            191.57 \\
\bottomrule
\end{tabular}

}
\caption{
Comparing mean file sizes for the reference systems and \SystemName{}.
Reference file sizes for \textit{Basic Median Cut} and \pngquant{} entries
have the same palette size as \SystemName{}.
}
\label{table:mean-absolute-file-sizes}
\end{table}

\section{Analyzing Compression Factors}\label{sec:validation}
We present an analysis of various factors affecting
the compression results for \SystemName{}. We explore the validity
of the $\delta_u$ functions constructed, the impact of the
multi-objective weighing factor, and the effect of varying a user's
color selection history length.

\subsection{$\deltau$: Data-Based Validation}\label{sec:validating-delta-u}
The large-scale vision data collected through the Specimen game can
enable novel post-hoc studies such as the re-quantization we explore
in this work.  At the same time, being data from a game rather than
a purposefully-designed user study, the Specimen dataset presents
an evaluation challenge when used in place of a user study.  We
present several analyses that address several of the challenges to
using the Specimen dataset and which provide us with greater
confidence in the validity of its use in studies such as ours.

\subsubsection{Correlation with Game Behavior}
As a first step towards a data-driven validation of our hypothesis
that potential artifacts produced by incremental quantization
are less likely to be perceived by a PCVD user, we
analyze the color changes performed by our approach in all 24
\pngquant pre-quantized Kodak images, when using the two
$\delta_u$ definitions. In order to isolate the effect of
$\delta_u$, we set the multi-objective weight ($\alpha$) to
1.0 such that we only perform color mergers as a function of
the scoring provided by $\delta_u$.

We consider each pixel in an incrementally quantized image to be
analogous to an observation in the Specimen
game. For example, given a pixel of color $c$, we say this is
equivalent to a Specimen game turn where the target background was
$c$.  We therefore consider pixels that are remapped to a different
color $c' \neq c$ as analogous to incorrect selections in Specimen.

For a given user, we took the Specimen game accuracies per hue-bucket,
and normalized these to [0.0, 1.0]. The normalization is done based
on the minimum and maximum values observed for that user, as follows:

\[ \frac{\mathrm{acc} - \Call{min}{\mathrm{acc}}}
  {\Call{max}{\mathrm{acc}} - \Call{min}{\mathrm{acc}}}, \]
where \textit{acc} corresponds to the user's accuracy for a particular hue-bucket.

\begin{figure}
\centering
\subfloat[Linear (230 colors)]{\includegraphics[width=0.475\columnwidth]{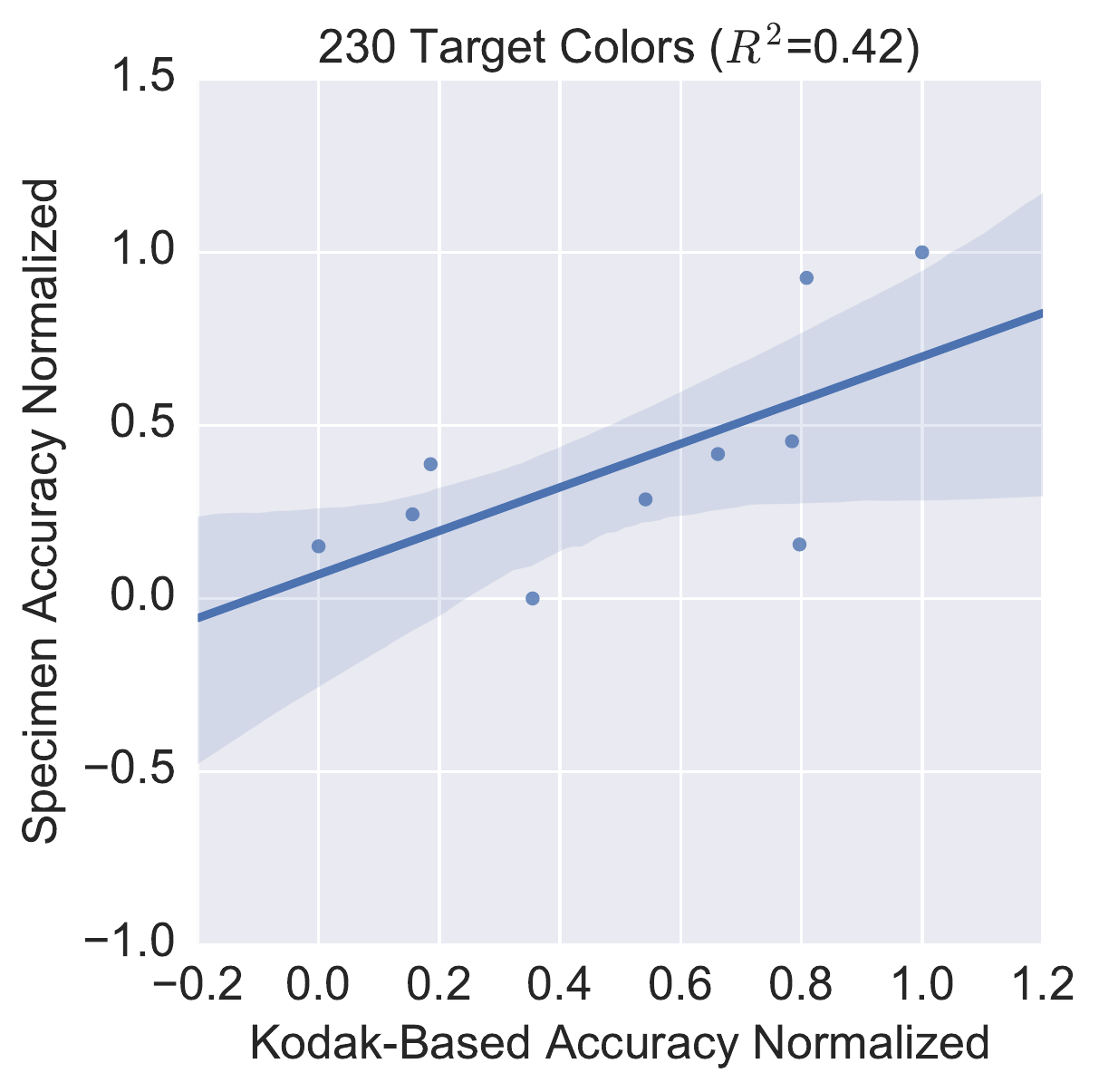}}~
\subfloat[Linear (204 colors)]{\includegraphics[width=0.475\columnwidth]{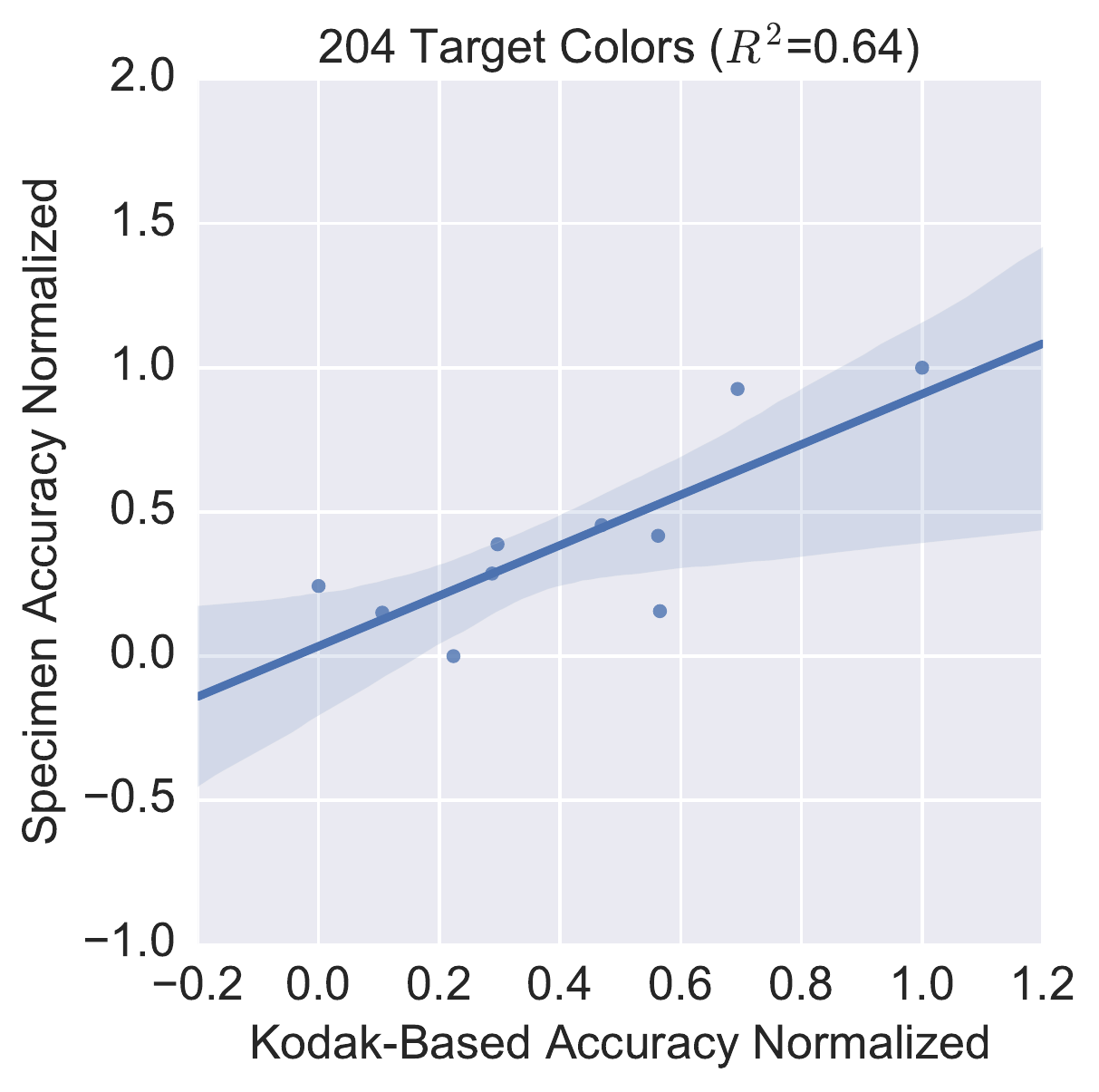}}\\
\subfloat[Linear (179 colors)]{\includegraphics[width=0.475\columnwidth]{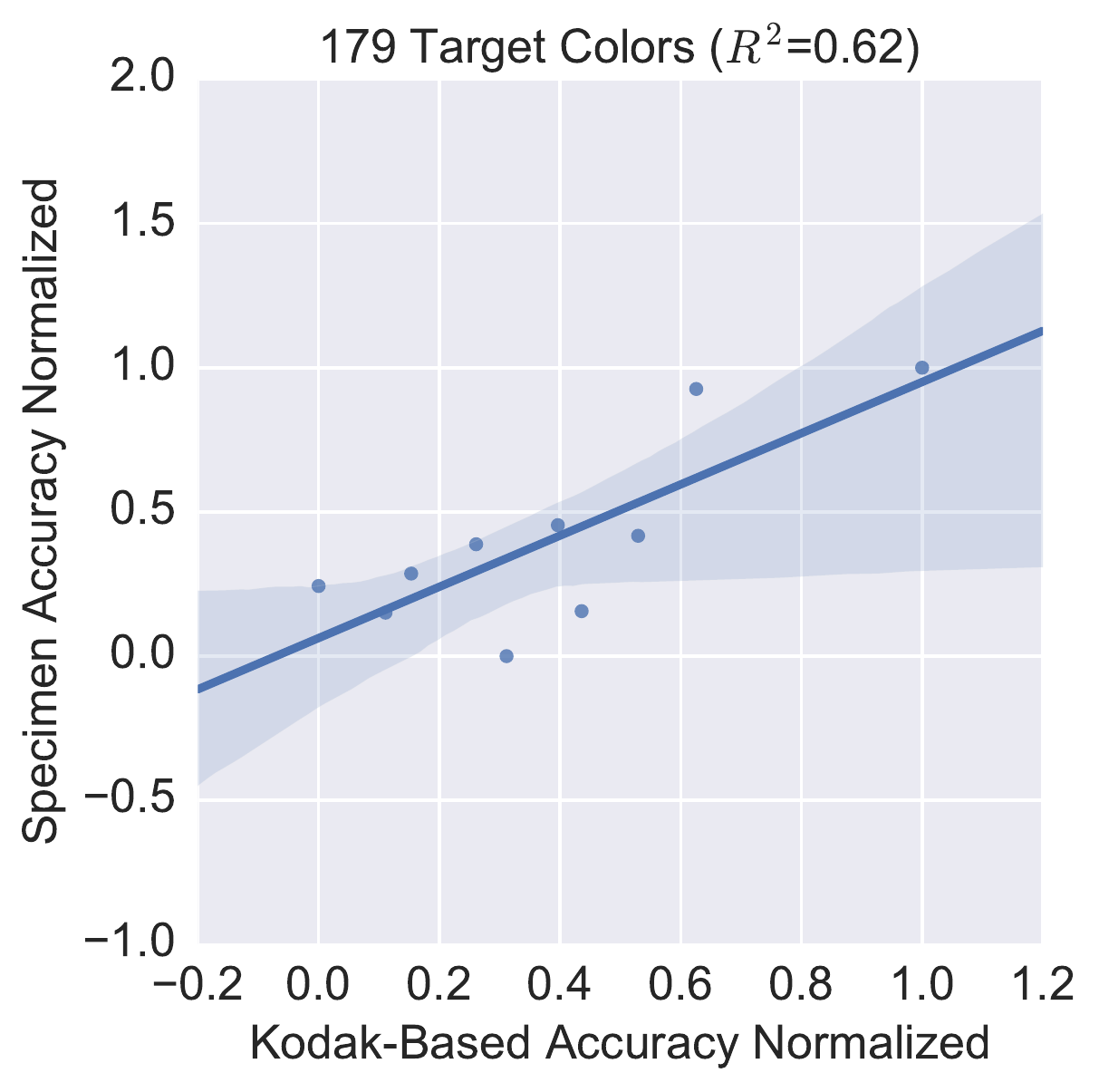}}~
\subfloat[Linear (153 colors)]{\includegraphics[width=0.475\columnwidth]{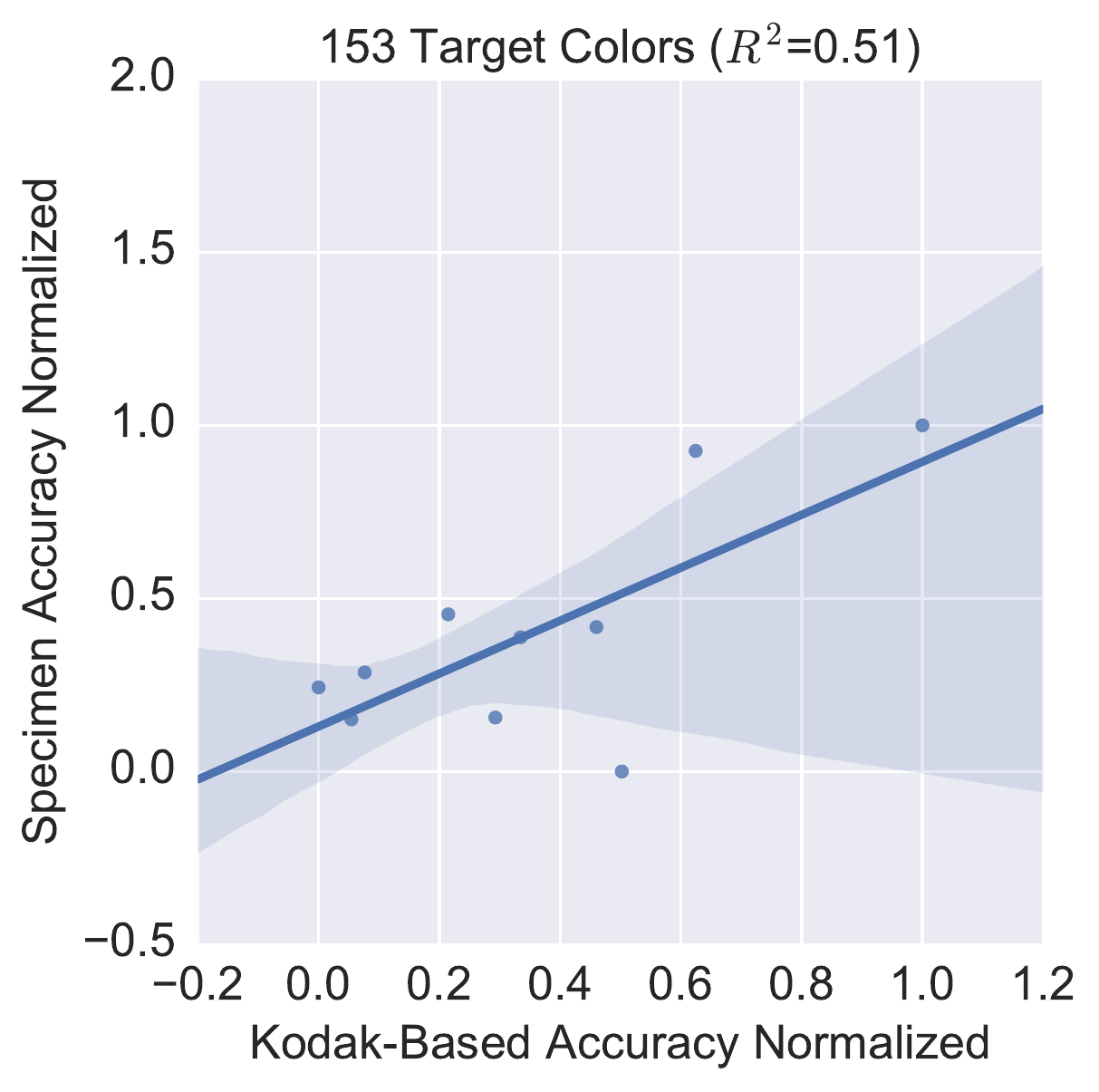}}\\
\subfloat[Linear (128 colors)]{\includegraphics[width=0.475\columnwidth]{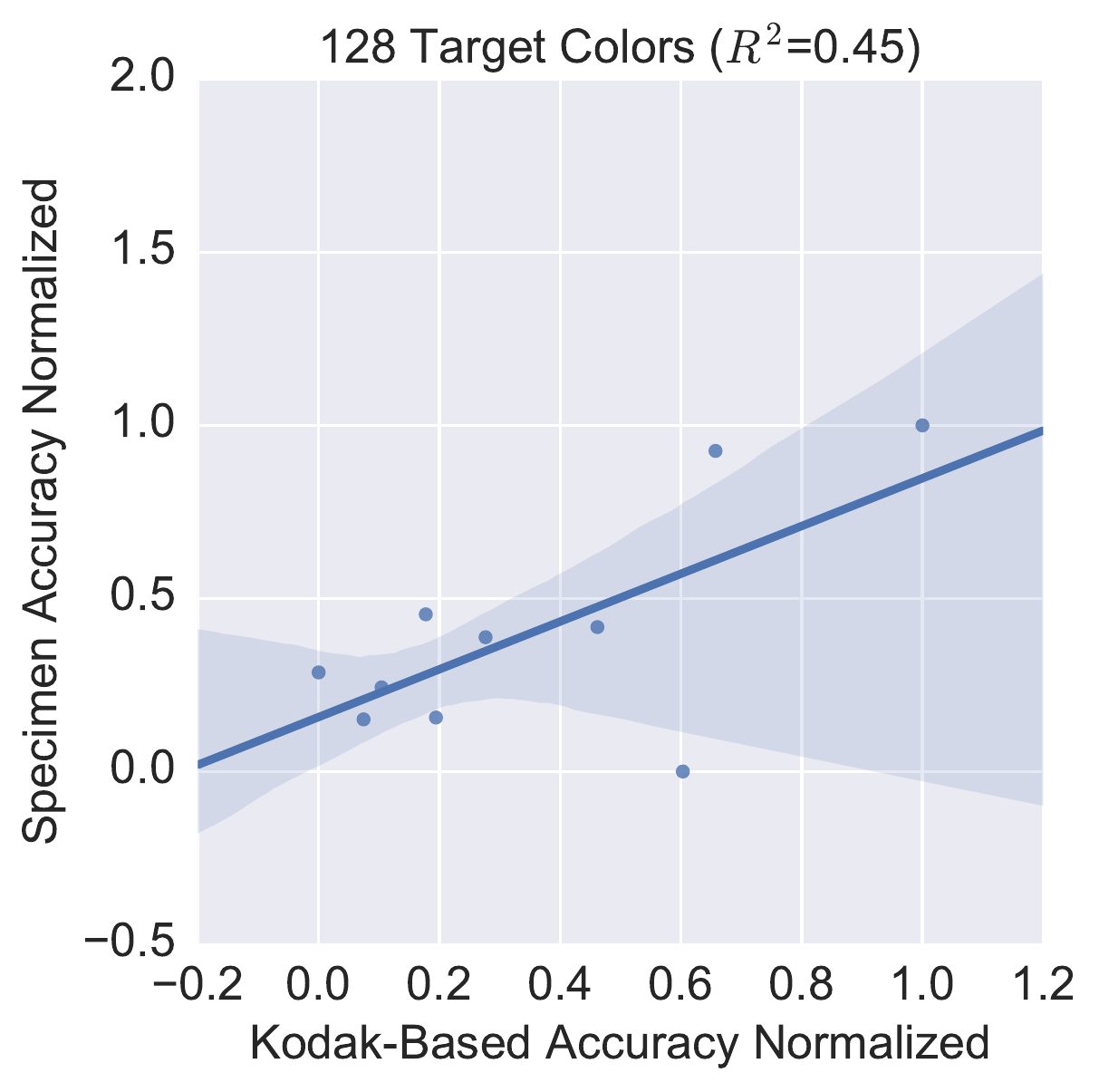}}\\
\caption{
We compare normalized hue-level accuracies for PCVD User 1 based on the
Specimen data and the recolorings of our benchmark images when using
the linear definition for $\delta_u$.  The line is a simple linear
regression and the shaded region corresponds to a 95\% confidence
interval.
}
\label{fig:simulated-hue-accuracies-linear}
\end{figure}

\begin{figure}
\centering
\subfloat[Non-Linear (230 colors)]{\includegraphics[width=0.475\columnwidth]{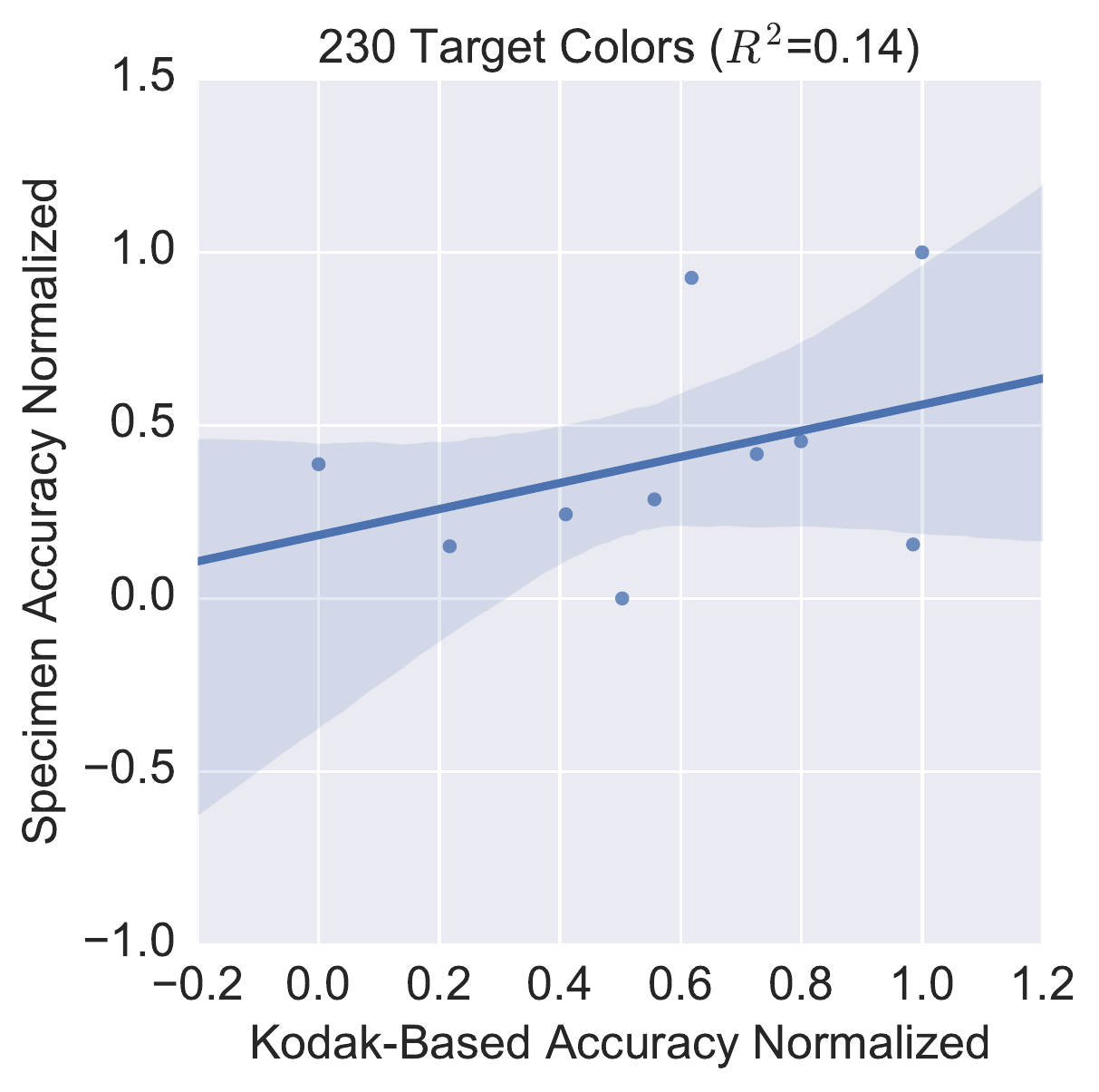}}~
\subfloat[Non-Linear (204 colors)]{\includegraphics[width=0.475\columnwidth]{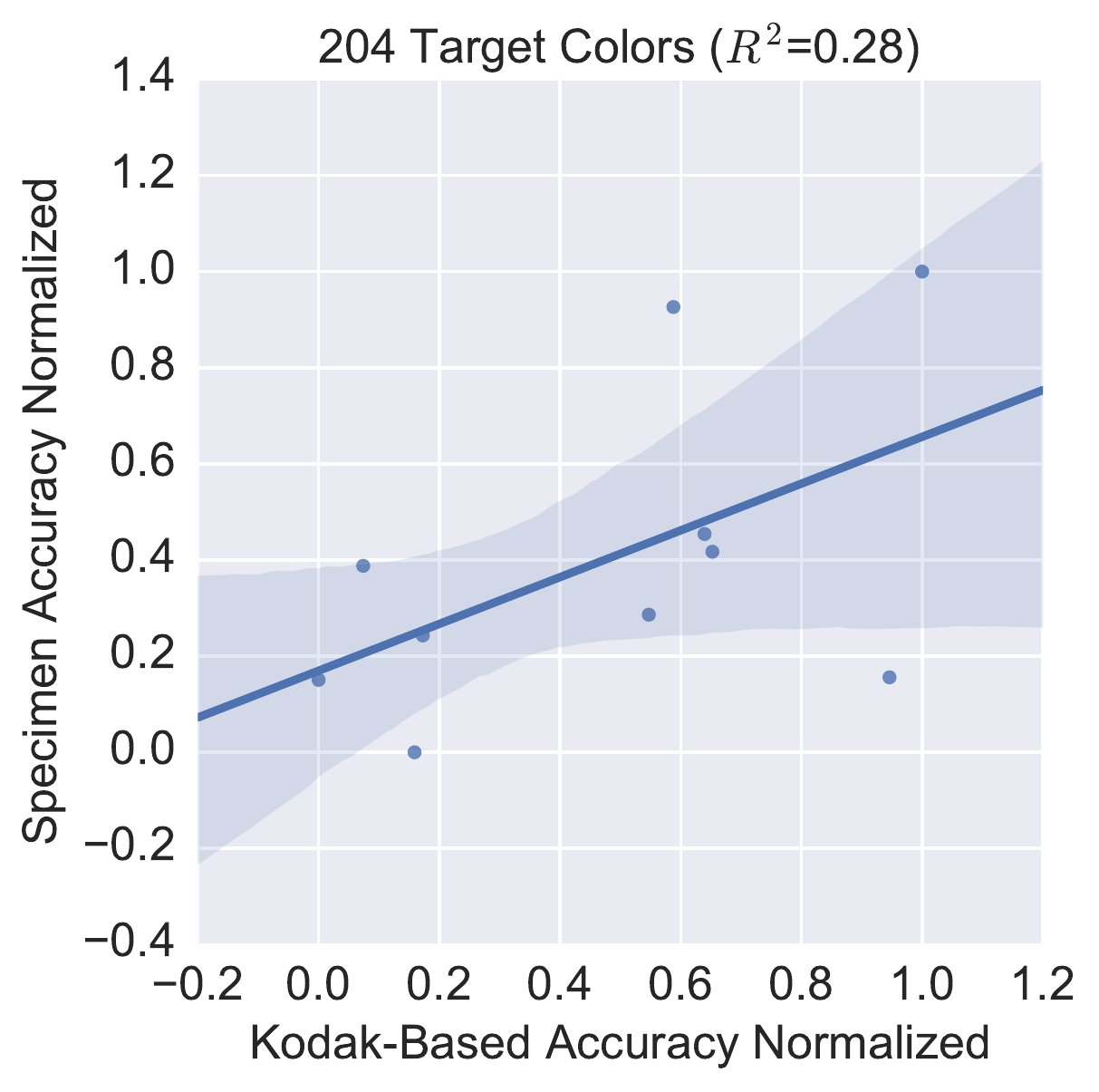}}\\
\subfloat[Non-Linear (179 colors)]{\includegraphics[width=0.475\columnwidth]{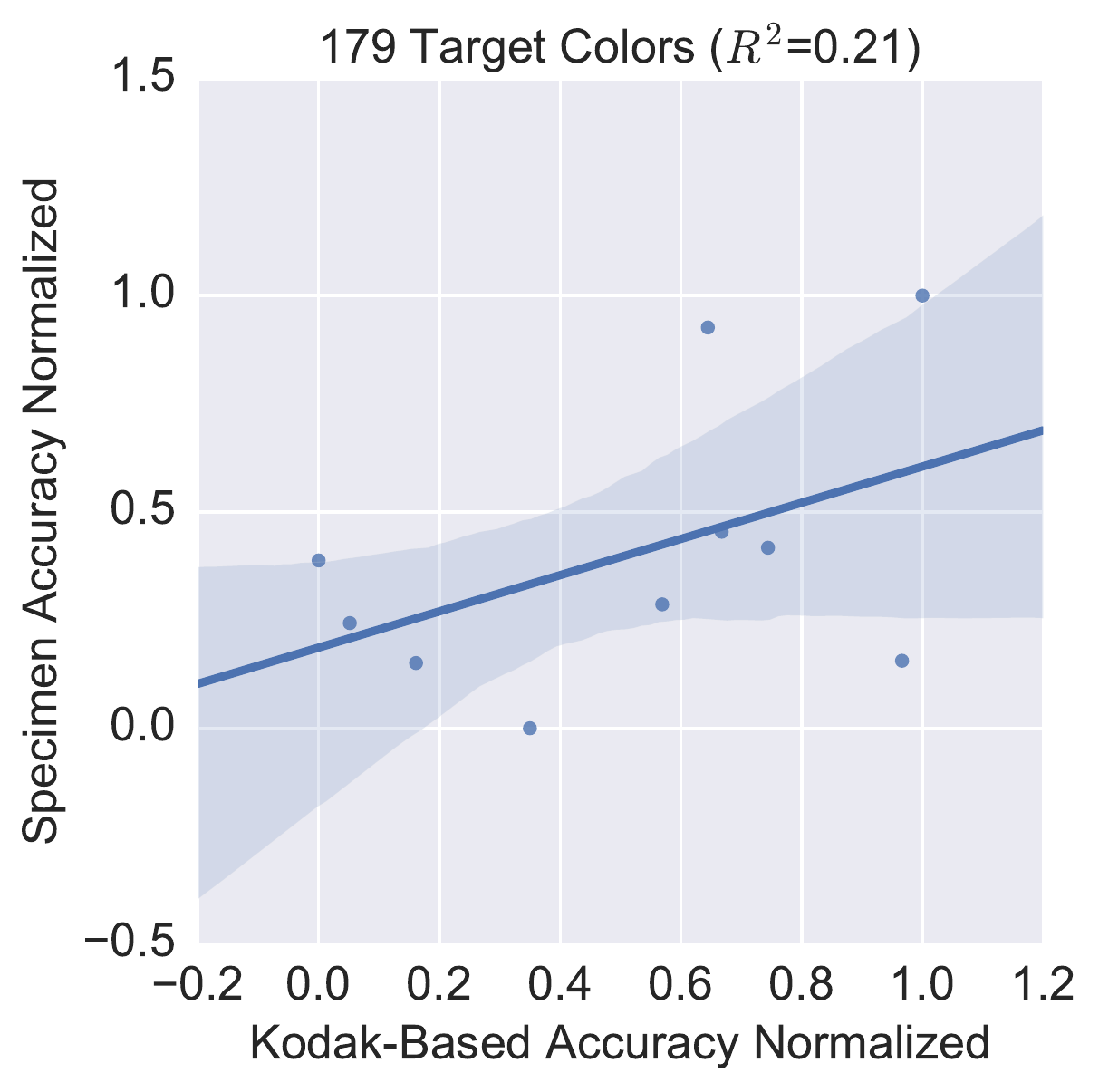}}~
\subfloat[Non-Linear (153 colors)]{\includegraphics[width=0.475\columnwidth]{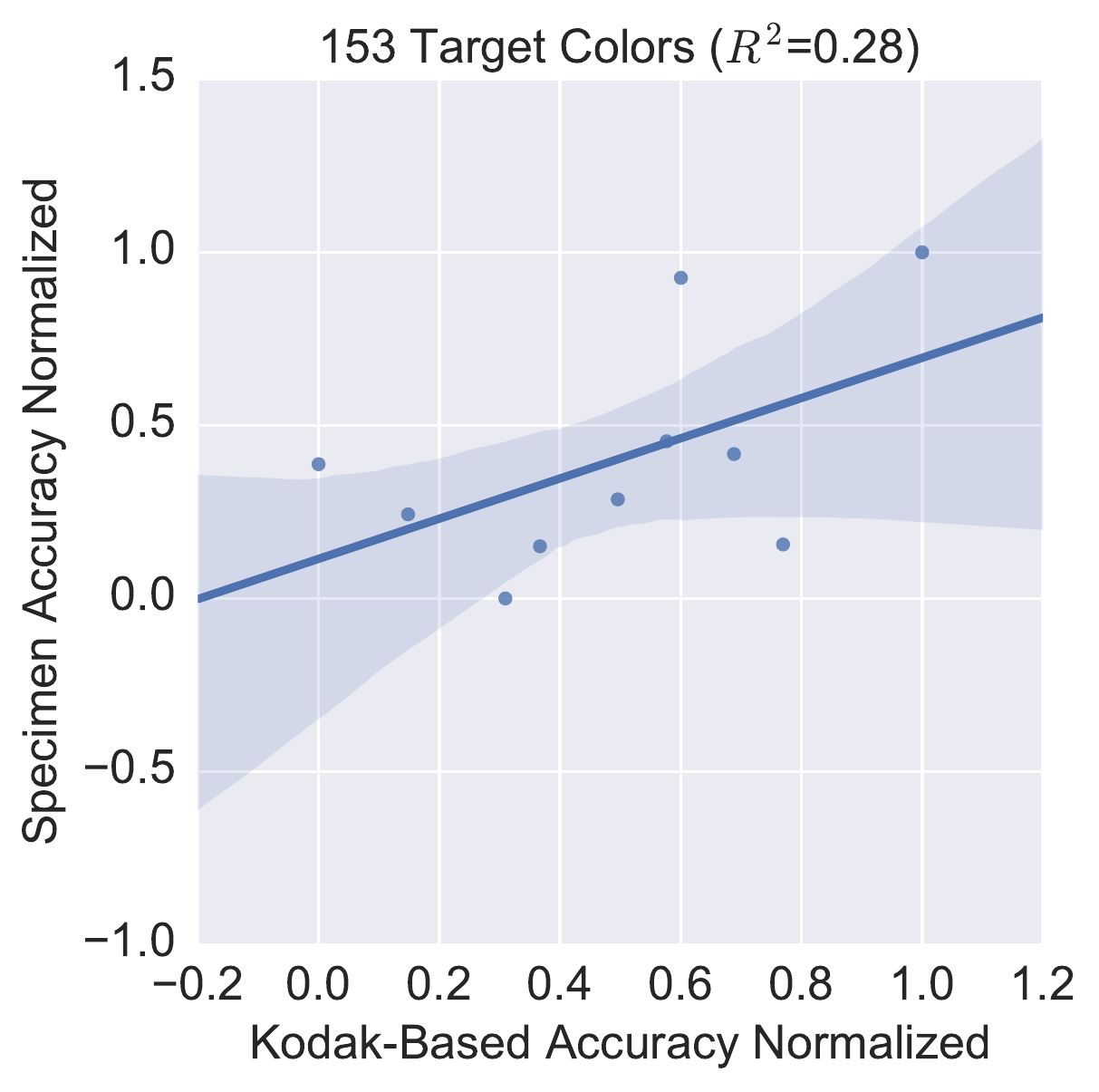}}\\
\subfloat[Non-Linear (128 colors)]{\includegraphics[width=0.475\columnwidth]{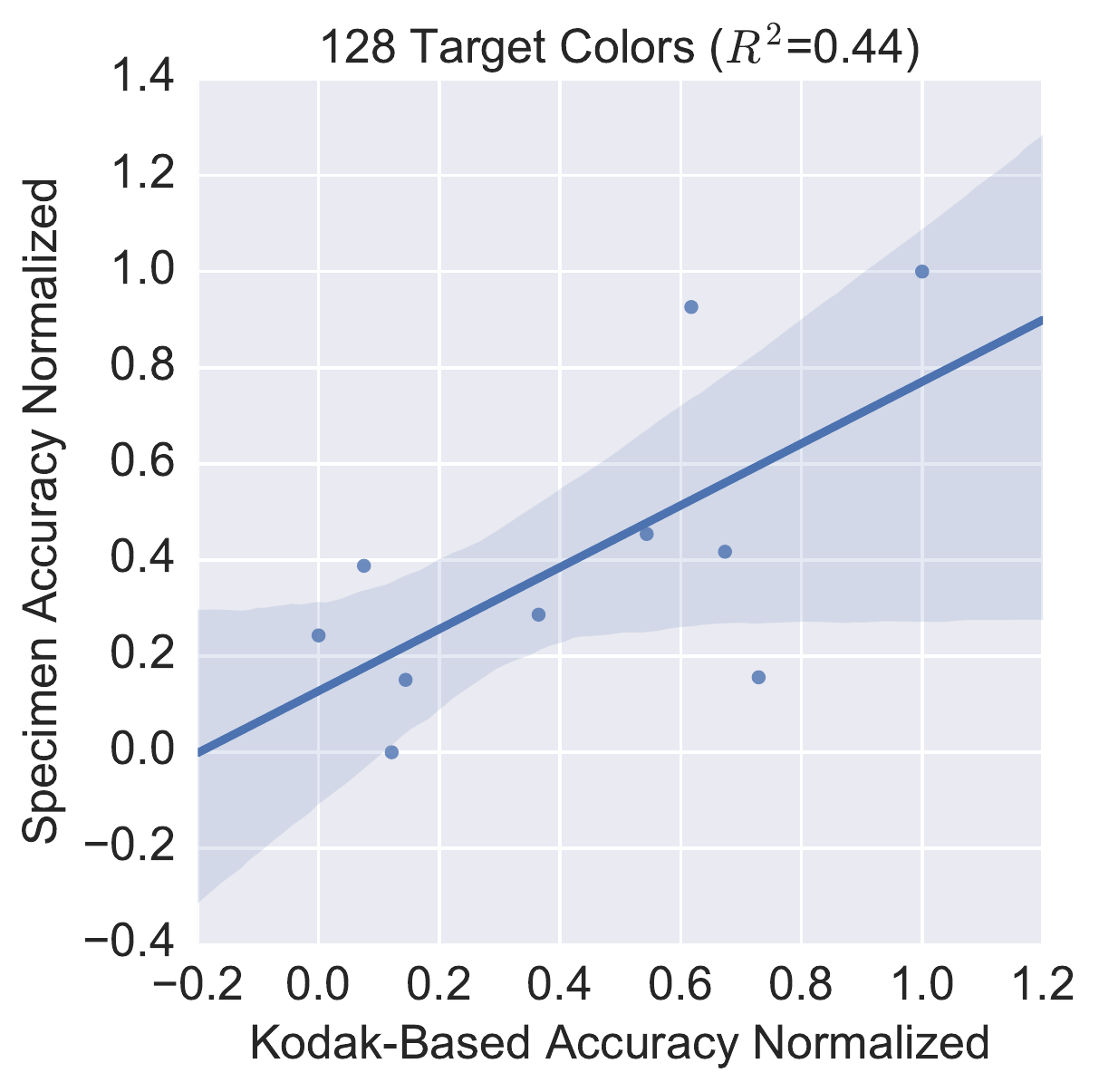}}\\
\caption{
We compare normalized hue-level accuracies for PCVD User 1 based on the
Specimen data and the recolorings of our benchmark images when using
the non-linear definition for $\delta_u$.  The line is a simple linear
regression and the shaded region corresponds to a 95\% confidence
interval.
}
\label{fig:simulated-hue-accuracies-non-linear}
\end{figure}

We then took the recolorings in our benchmark image set, aggregated
them across images, computed hue-bucket accuracies and performed a
similar normalization. We refer to these values as the \textit{Kodak-based accuracies}.
Figure~\ref{fig:simulated-hue-accuracies-linear}
and Figure~\ref{fig:simulated-hue-accuracies-non-linear} show the
results over the different palette sizes in our experiment when using the linear and non-linear versions
of $\delta_u$, respectively. Each plot shows the Kodak accuracies
on the x axis, and the Specimen accuracies on the y axis (points)
together with a linear regression through these points (line). The
$R^2$ value for this regression is given in the title of each plot.
The shaded region around the regression line in the plot is the
95\% confidence interval for the regression line. When considering
a given image as opposed to the aggregated results, accuracies shift
lower for images when more aggressively compressing (lower target
number of colors).  The accuracies observed across palette sizes
have good correlation with observed hue-level accuracies for PCVD
User 1 in Specimen with correlation coefficients between
\lowHueAccuracyCorrelationLinear{} and \highHueAccuracyCorrelationLinear{}
for the linear $\delta_u$ and \lowHueAccuracyCorrelationNonLinear{}
and \highHueAccuracyCorrelationNonLinear{} for the non-linear
variant.

\subsubsection{Reducing Distances in Transformed Space}
We do not claim that our linear and non-linear transformations will
fully predict the mapping from observed target colors to the set
of colors perceived by the user.  However, our definition of
$\delta_u$ does not require prediction of pairs of colors, but
rather the ranking of colors based on perceived color
equivalence.

With this goal in mind, we evaluate the transformations
by comparing the change in distance between colors in the transformed
user space relative to the original CIE-LAB color space. For each user,
we take all the color confusions observed in the Specimen data and 
split the data into a training and test set of observations.
We remove from the test set any pair of colors in which either the 
target or the specimen color was observed in the training set.
We then fit the transformation using the training set
and evaluate (using the test set) the change in distance from
the original color space to the transformed color space. This process
is repeated with 10 different splits of the underlying data.

Table~\ref{table:distance-changes} shows the results for the top three
ranked PCVD users in our evaluation set of 30 users. Both the linear
and non-linear transformation successfully reduce the distance despite
not observing the exact pairs of colors from the test data set.
The reductions in distance ranged from \lowDistanceReductionPct{}
to \highDistanceReductionPct{}.

\begin{table}
\resizebox{\columnwidth}{!}{
\begin{tabular}{lll}
\toprule
PCVD User & Transformation & Mean Change Distance (+/- Std) \\
\midrule
        1 &     Non-Linear &                -0.65 (+/-0.10) \\
        1 &         Linear &                -0.54 (+/-0.11) \\
        2 &     Non-Linear &                -0.80 (+/-0.08) \\
        2 &         Linear &                -0.73 (+/-0.12) \\
        3 &     Non-Linear &                -0.64 (+/-0.08) \\
        3 &         Linear &                -0.54 (+/-0.11) \\
\bottomrule
\end{tabular}

}
\caption{
The average change in distance between colors in the transformed
color space compared to the original LAB color space for the top
three PCVD users, as ranked by our heuristic $\mathcal{H}$.  The
mean change is computed over a held-out portion of the data set.
Values are computed over 10 different splits of the data.
}
\label{table:distance-changes}
\end{table}

\subsection{$\alpha$: Evaluation of the Multi-Objective Weighting Factor}
The multi-objective weighting factor, $\alpha$ (Equation~\ref{eqn:combinedGoal} and
Equation~\ref{eqn:optimizationGoal}), can impact the file
size produced by our incremental quantization technique.  Larger
values of $\alpha$ favor the color equivalence measure produced
by $\delta_u$ over the number of pixels associated with a particular
color when selecting the color pairs to merge in the new palette.
We present results for PCVD User 1, as their color selection history
produced the highest heuristic score for PCVD identification (see
Section~\ref{sec:pcvd} for details) and so we consider them a good example of the
color vision deficient individual that could benefit from \SystemName{}.

\begin{figure}
\centering
\includegraphics[width=0.49\textwidth]{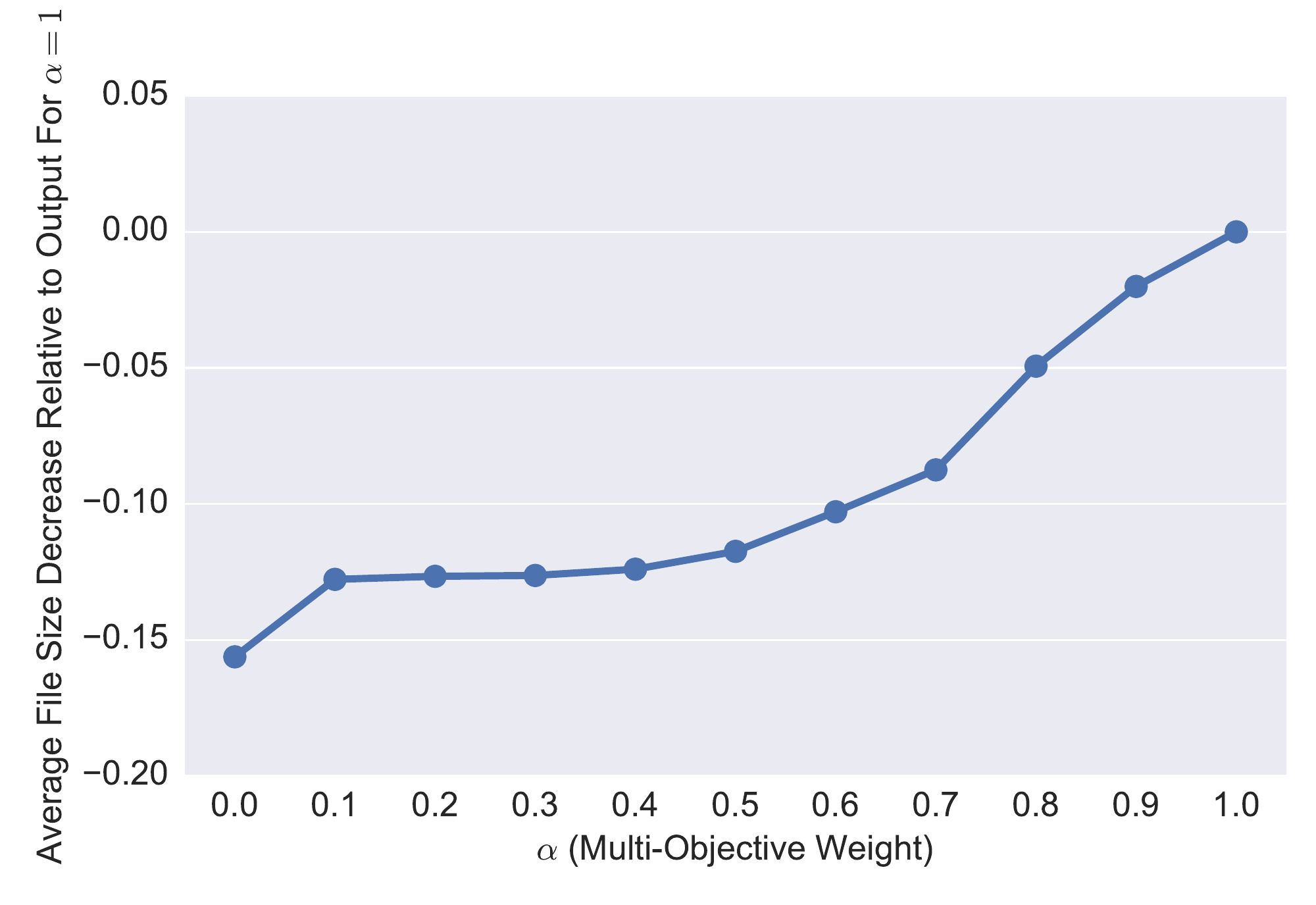}
\caption{
Average file size reduction for varying values of $\alpha$, the
multi-objective weighting factor, across the 24 Kodak images.  File
size reductions are relative to the file size when $\alpha=1$.
Evaluation done with a target palette size of 204 colors using PCVD
user 1's non-linear transform-based $\delta_u$. Lower values of
$\alpha$ produce, as expected, smaller files.
}
\label{fig:multi-objective-comparison}
\end{figure}

Figure~\ref{fig:multi-objective-comparison} shows the average file size
reduction for Kodak PC Set images when incrementally quantized using
the non-linear transform-based $\delta_u$ for PCVD User 1 with a
target palette size of 204 colors. We compute the file size reduction
relative to the file size for the same image when $\alpha=1$.
We see that the average file size decreases as expected when we
reduce the value of $\alpha$. The output file size at $\alpha=0$
is approximately 15\% smaller than the output file size at $\alpha=1$,
on average across the Kodak benchmark set for PCVD User 1.

\subsection{History Impact: Handling New Users}\label{sec:history}
The length and variety of a user's Specimen history is important
for the construction of $\delta_u$.
Given that Specimen is a game, some users may play it more
frequently than others, producing histories of differing length
and with different target and specimen color observations.

Users with longer histories of
color selections have a better chance of observing a larger variety
of colors within the Specimen game and will
have more observations across confusions to better estimate the
appropriate transformation for $\delta_u$. 
So a natural question is: \textit{How can we construct $\delta_u$ for
a new user?} As an initial step in this direction, we consider a
related question: \textit{How do limited histories affect $\delta_u$?}

With limited user histories, the impact on compression is driven
by the extent to which the $\delta_u$ constructed from shorter
histories produces a sorted list of color merger candidates in an
image that differs significantly from the list that would be produced
under the $\delta_u$ constructed from the complete user history.
This effect is dependent on both $\delta_u$ and the specific image
to be compressed, as well as the multi-objective weighting factor
$\alpha$.  To evaluate this potential impact, we consider the 24
Kodak images from our evaluation set and compared the changes in
sorted merger candidates. For all comparisons we set $\alpha=1$ to
isolate the impact of the change in the user history.  For the
non-linear $\delta_u$ we initialize the random state
for model fitting to be the same value in all experiments to
consistently compare against the same original sorted candidate
list. We take the top 200 color merger candidates, as ranked by the
$\delta_u$ with full history, and measure the rank of these candidates
in the new sorted merger candidate list produced by the limited
history $\delta_u$. We then compute Spearman's rho over the two
sets of rankings.  Spearman's rho is a common non-parametric measure
for rank correlation~\cite{spearman1904proof}.

\begin{figure}
\includegraphics[width=\columnwidth]{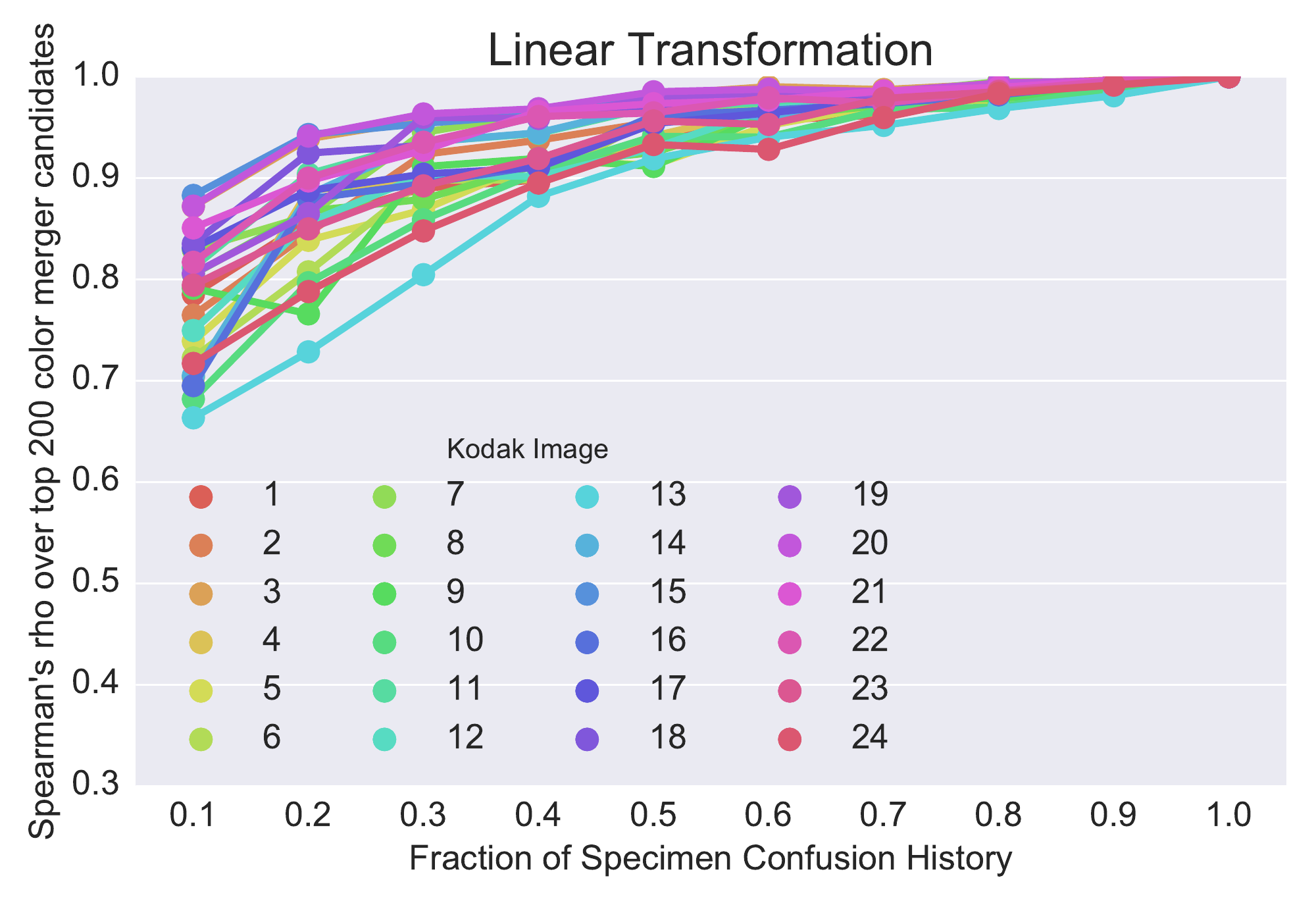}
  \caption{
  Rank correlation for color merger candidate lists produced for
  PCVD User 1 when using the linear $\delta_u$s constructed from the full
  history and a limited fraction of their history.
  }
  \label{fig:limit-history-rho-linear}
\end{figure}

\begin{figure}
\includegraphics[width=\columnwidth]{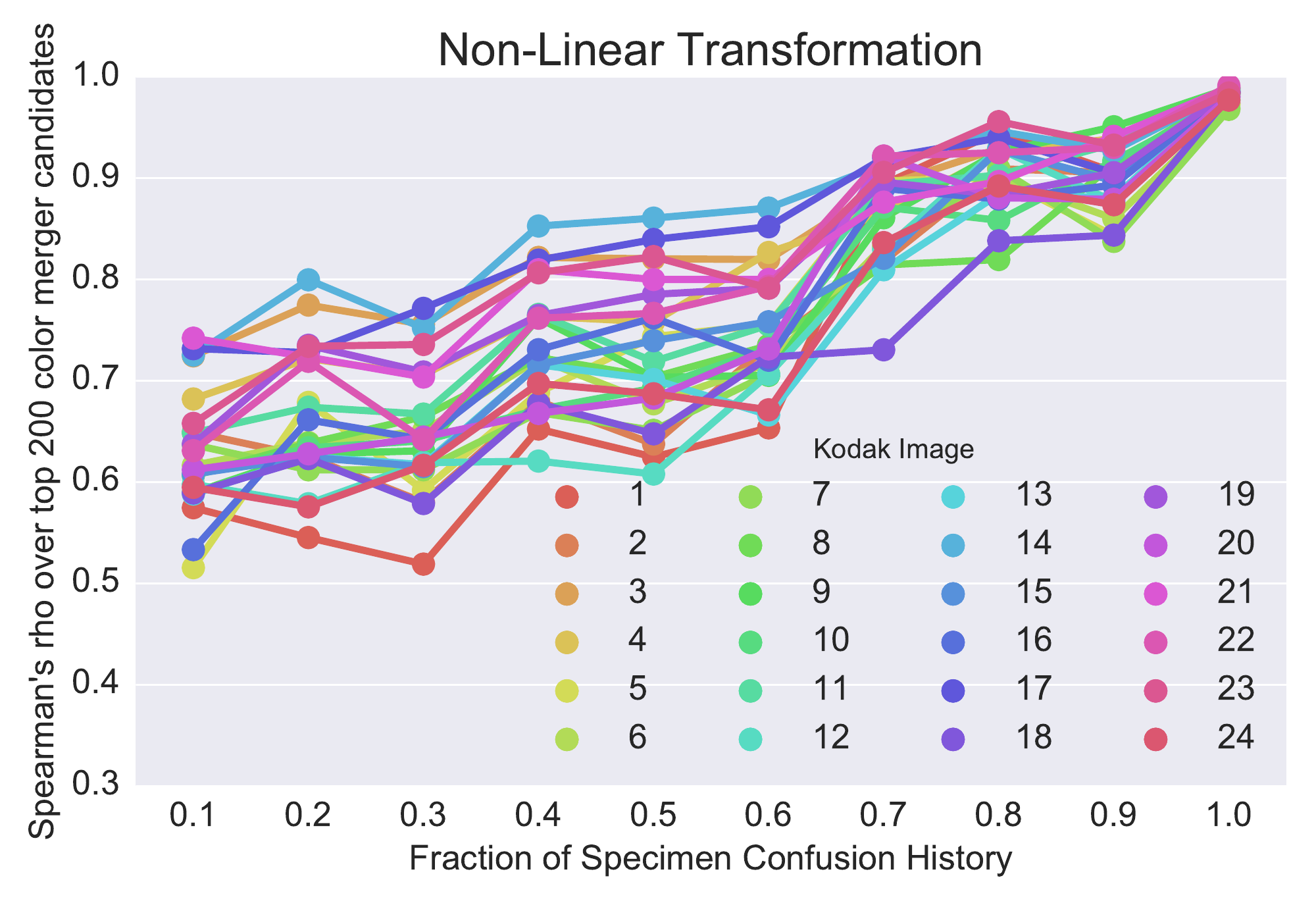}
  \caption{
  Rank correlation for color merger candidate lists produced for
  PCVD User 1 when using the non-linear $\delta_u$s constructed from the full
  history and a limited fraction of their history. More complex
  transformations, such as the non-linear transform, are more heavily
  impacted by limited histories.
  }
  \label{fig:limit-history-rho-non-linear}
\end{figure}

Figure~\ref{fig:limit-history-rho-linear} and
Figure~\ref{fig:limit-history-rho-non-linear} shows the Spearman's rho results for
PCVD User 1. The changes in ranking correlation are not uniform
across images, as the candidate list produced directly depends
on the original image palette.  More complex transformations, such
as the non-linear transformation, produce lower rank correlations
with limited history. One simple solution for addressing this effect
of limited user histories is to employ a simpler $\delta_u$ until
the user accumulates enough confusion observations to shift to a
more sophisticated $\delta_u$ implementation.

\section{Related Work}
Hu et al.~\cite{hu2016multiobjective} propose to address the
quantization decision as a multi-objective optimization problem,
which they solve using an evolutionary approach. While we also
propose a multi-objective optimization formulation of incremental
quantization, our objectives differ from theirs. We aim to maximize
the score of the quantization map as a function of existing
user-specific confusions and the number of pixels associated with
a color in the original image. Their objective function aims to
minimize the distance of the original colors to the quantized palette
and maximize the distance across colors mapped to different elements
in the output palette.

Using a swarm-based optimization algorithm, 
Kubota et al.~\cite{kubota2015color} quantize an image's color palette and apply dithering to reduce
the perceived degradation in image quality. Their technique
accounts for human perception exclusively by considering
a weighted distance between the original and quantized images.
In contrast to their approach, we account for human perception
by empirically constructing a function that effectively
ranks colors that a user on a mobile display repeatedly 
confused while playing a game focused on color comparisons.

In order to produce re-colorings of images with fewer colors, 
Rasche et al.~\cite{rasche2005re} explore a technique that preserves
image details. Their target recoloring aims to preserve contrasts in
the original image and maintain luminance consistency. They extend
this technique to recolor images for observers with color vision
deficiencies. In contrast, \SystemName{}'s goal is not to correct
images for improved viewing by color-vision-deficient users, but
rather to exploit their perceptual differences to produce a more aggressive
compression that provides access to improved storage and network communication. Similar to this work, we start with a
quantized image, which allows our approach to scale to standard
images with large palettes and to preserve
compression benefits for color-vision-normal individuals when
reverting incremental compression.

Perceptual image compression aims to improve compression ratios
while minimizing the impact on the perceived image
quality~\cite{eckert1998perceptual, nadenau2000human}.  Recent work
~\cite{yao2017novel} proposed a new image compression algorithm
based on contrast sensitivity in the human visual system. In contrast
to our approach, existing work has not targeted color-vision-deficient viewers.

Existing work in computation for color blind users has focused on
accessibility. The authors in~\cite{jefferson2007interface} present
a system to adapt a digital image for a color-vision-deficient
individual. They show that their transformation reduces the error
in a user study using Ishihara plates~\cite{ishihara1960tests}.
The interface presented relies on a color vision model
that transfers stimulus across RGB channels based on the type
of color vision deficiency: protanopia, deuteranopia, and tritanopia.
\SystemName{} instead relies on a history of color comparisons (collected
through a game for our implementation) to build a bespoke
quantification of color equivalence.

The technique presented in~\cite{iaccarino2006efficient} adapts
website content for color-vision-deficient viewers. Their algorithm
is customizable and opens up personalized transformations. Similarly,
our work provides a technique that can be customized for a given
user. In contrast, our compression algorithm does not focus on
accessibility but instead on exploiting color vision deficiencies
for more efficient image representation.

Stanley-Marbell et al.~\cite{stanley2016crayon} exploit the inherent approximation in
human visual perception to transform image shapes and colors to
improve power dissipation on OLED displays. We similarly exploit
the human visual system, but focus on a subset of the population
(color-vision-deficient users) and develop a novel approach that
can reduce storage and improve client communication for remote image
services or mobile devices.

\section{Conclusion}
This paper provides results for incremental quantization targeting
color-vision-deficient users. We introduced a user-specific function
($\delta_u$), empirically constructed from mobile game data, that
quantifies perceived color equivalence for color-vision-deficient individuals.
We evaluated two possible implementations of $\delta_u$ based on a
dataset of \SizeOfDataset{} color comparisons collected through a
mobile game. We produced implementations for 30 distinct users in
our data and showed that we can achieve incremental compression
over state-of-the-art methods.

\SystemName{}, an implementation of our incremental quantization
algorithm, on average reduced file sizes in our benchmark by
\avgLowPercent{} to \avgHighPercent{} over outputs with the same
palette size produced by popular compressors.  Our quantization
algorithm also produces a simple mapping of quantization decisions
that recovers the original image, which allows standard viewing by
color-vision-normal individuals.

Our analysis shows that our $\delta_u$ definitions correlate
well with users' game history and that the transformations they are
based on successfully reduce the distance ratio for colors in the
transformed space relative to the original color space. We explored
the impact that limited user history has on the outputs of $\delta_u$
and show that the impact of limited history changes depending on
the user and the image being quantized.

\acknow{We thank Erica Gorochow, Salvatore Randazzo, and Charlie Whitney for providing access to the Specimen dataset. PSM is supported by an Alan Turing Institute award TU/B/000096 under EPSRC grant EP/N510129/1.}

\showacknow 

\pnasbreak

\bibliography{daltonquant}

\end{document}